\documentclass[12pt]{article}

\usepackage{newtxtext,newtxmath}
\usepackage{hyperref}

\usepackage{booktabs} 
\usepackage{siunitx} 
\usepackage{graphicx}
\usepackage{textcomp}
\usepackage{afterpage}
\usepackage{multirow,hhline}
\usepackage{siunitx}
\usepackage{enumerate}  
\usepackage{comment}
\usepackage[left]{lineno} 
\usepackage[letterpaper,margin=1in]{geometry}
\usepackage{gensymb}
\usepackage{siunitx}
\usepackage{xcolor} 

\usepackage{caption}
\usepackage{setspace}
\usepackage{newfloat} 
\renewenvironment{abstract}
	{\quotation}
	{\endquotation}

\date{}

\makeatletter
\renewcommand{\fnum@figure}{\textbf{Figure \thefigure}}
\renewcommand{\fnum@table}{\textbf{Table \thetable}}
\makeatother

\usepackage{cite}

\usepackage{url}

\usepackage[utf8]{inputenc} 
\DeclareUnicodeCharacter{2032}{$'$} 

\def\scititle{
Kinetics of Vacancy-Assisted
Reversible Phase Transition in Monolayer MoTe$_2$
}
\title{\bfseries \boldmath \scititle}

\author{
	Fei Shuang$^{3,4}$,
	Daniel Ocampo$^{1}$,
    Reza Namakian$^{1}$, Arman Ghasemi$^{4}$\and
	Poulumi Dey$^{3}$,
    Wei Gao$^{1,2,4\ast}$\and
\begin{tabular}{@{}p{\textwidth}@{}}
    \raggedright 
    \small $^{1}$J. Mike Walker’66 Department of Mechanical Engineering, Texas A\&M University, College Station, TX 77843, United States.\\
    \small $^{2}$Department of Materials Science \& Engineering, Texas A\&M University, College Station, TX 77843, United State.\\
    \small $^{3}$Department of Materials Science and Engineering, Faculty of Mechanical Engineering, Delft University of Technology, Mekelweg 2, Delft, 2628 CD, The Netherlands.\\
    \small $^{4}$Department of Mechanical Engineering, The University of Texas at San Antonio, San Antonio, Texas 78249, United States.\\
    \small$^\ast$Corresponding authors. Emails: wei.gao@tamu.edu
\end{tabular}
}

\begin{document} 
\maketitle

\newpage
\begin{abstract} \bfseries \boldmath

We investigate the kinetics of phase transition between the 2H and 1T$^\prime$ phases in monolayer MoTe$_2$ using atomistic simulations based on a machine learning interatomic potential trained on SCAN-DFT data, combined with mean field kinetic theory to interpret the underlying mechanisms. The transition is found to involve both diffusive and diffusionless mechanisms. Nucleation of 1T$^\prime$ phase is initiated by the coalescence of neighboring Te monovacancies into divacancies, which are found to be mobile and can interact with other Te vacancies to form small triangular 1T$^\prime$ islands. Growth of these islands proceeds either by incorporating pre-existing vacancies at the phase boundaries or, in their absence, by absorbing divacancies that migrate from the surrounding lattice. Once a critical island size is reached, vacancy-free growth becomes possible although with a higher activation barrier. Upon removal of external stimuli, the system reverts to 2H phase, during which Te vacancies reorganize into three-fold spoke-like vacancy lines at the island center. This reverse process and the subsequent 1T$^\prime$$\leftrightarrow$2H reversible transitions are diffusionless, rapid, do not require additional vacancies and can be driven by mild external stimuli. Although our analysis focuses on strain-induced transitions, the kinetic mechanisms are expected to be generalizable to other types of stimuli. 
\end{abstract}

\newpage

\section*{Introduction}

\noindent

Phase transitions of two-dimensional transition metal dichalcogenides (2D TMDs) have attracted significant research attention in recent years due to their promising applications in atomically thin devices, including memory devices \cite{wuttig2007}, reconfigurable circuits \cite{wang2016optically} and topological transistors \cite{qian2014quantum}. However, the kinetic mechanisms governing these transitions still remain poorly understood. Among 2D TMDs, MoTe$_2$ is notable due to the small potential energy difference between its semiconducting 2H and conducting 1T$^\prime$ phases\cite{Duerloo2014}, making the phase transition achievable under moderate conditions\cite{Li2021}. Phase engineering in MoTe$_2$ can be realized  during synthesis, for instance, via chemical vapor deposition (CVD), by tuning parameters such as Te feed rate\cite{Yoo2017}, metal precursor selection\cite{Zhou2013}, substrate choice\cite{Hynek2021}, and synthesis temperature\cite{Keum2015,Sung2017}. Alternatively, phase transitions in 2D TMDs have been induced through a variety of external stimuli.

The external stimuli used in experiments to induce phase transition in 2D TMDs can be broadly categorized into three groups. The first category includes methods like alloying\cite{Duerloo2016,Zou2018,Deng2021} and ion intercalation with alkali metals\cite{Han2021}, which drive phase transitions by introducing foreign elements. These approaches may compromise the intrinsic properties of the original TMDs. The second category involves direct energy inputs, such as electron beam and laser irradiation\cite{Lin2014,Tan2018,Si2019,Ryu2023,Nanotechnology2024,Cho2015}. These may result in irreversible transitions\cite{Cho2015} or degradation of the material’s structural integrity and functionality\cite{Sun2019,Akinwande2017}. The third category include less intrusive methods, such as mechanical strain\cite{Duerloo2014,Song2016}, electron doping\cite{Kang2014,Kim2017}, electrostatic gating\cite{Li2016-eg,Wang2017}, and combinations of these techniques, for example, strain coupled with gating\cite{Awate2023}. These methods offer more precise and reversible control over phase transitions.

Previous experiments have shown that Te vacancies play a critical role in the phase transition of MoTe$_2$\cite{Lin2016,Zhou2021,JPCC2021,Wang2023,Kim2024,Okello2024}, as they help stabilize the 1T$^\prime$ phase by lowering its potential energy. For example, Te vacancy formation is a key factor in irradiation-induced phase transitions\cite{Cho2015}, where a high density of Te vacancies drives the 2H$\rightarrow$1T$^\prime$ transformation. However, under such high vacancy concentrations, the transition becomes irreversible, as the system settles into a lower-energy final state. By contrast, reversible phase transitions are more desirable for semiconductor device applications. Therefore, our current study focuses on reversible 2H~$\leftrightarrow$~1T$^\prime$ transitions under low vacancy density, either intrinsically present in materials or introduced through targeted defect engineering. We demonstrate that the phase transition pathways depend on the spatial distribution and mobility of vacancies, and are governed by the applied external stimuli.

On the theoretical and computational side, density functional theory (DFT) has been used to study the thermodynamics and kinetics of phase transitions in 2D MoTe$_2$. It has been shown that external stimuli can lower the energy of the 1T′ phase, making the 2H $\rightarrow$ 1T$^\prime$ transition thermodynamically favorable. In terms of kinetics, most studies focused on calculating the transition energy barrier within a unit cell. Reported barriers can be reduced below 0.9 eV per formula unit\cite{Duerloo2014,Ghasemi2020,krishnamoorthy2018,Yu2023}, and under excited state, the barrier can drop further to as low as 0.08 eV per formula unit\cite{krishnamoorthy2018}. Although these barrier values appear small, they do not necessarily imply that the phase transition is readily thermally activated. Because the phase transition calculated in this way occurs in a concerted manner involving many formula units, the total energy barrier scales with the number of units, leading to a significantly larger effective barrier. Therefore, energy barriers calculated based solely on a unit cell do not capture the full complexity of the phase transition kinetics in MoTe$_2$.

DFT calculations have been used to investigate the role of Te vacancies in the kinetics of phase transitions in MoTe$_2$\cite{JPCC2021,Si2019}. It has been shown that the migration energy barrier of Te monovacancies is significantly lowered in the excited states, playing a critical role in the photoinduced phase transition of MoTe$_2$\cite{Si2019}. This suggests a diffusive mechanism driven by Te monovacancy migration. However, in the absence of excitation, the Te monovacancy migration barrier remains high, around 1.6 eV\cite{JPCC2021}, suppressing the diffusion under ambient conditions. In addition, Te vacancy lines in the 2H phase have been shown to facilitate fast and reversible phase transitions\cite{JPCC2021}, although the underlying mechanisms of vacancy line formation remain unclear.

One of the key challenges in modeling phase transitions in MoTe$_2$ is the limited spatial and temporal scales accessible by DFT, which constrain its ability to resolve atomistic kinetic mechanisms. Existing
empirical potentials such as the Stillinger-Weber potential are not adequately developed to handle defects and transitions between two phases\cite{Jiang2017}. 
Recently, machine learning interatomic potentials (MLIPs) have emerged as powerful tools for studying atomistic mechanisms in solid deformation \cite{Yin2021,Mortazavi2021} and phase transitions \cite{Jinnouchi2019,Zhou2023-1,Liu2024} with near-DFT accuracy. In this study, we developed an accurate MLIP, which is specifically trained to capture complex defect behavior and its interactions with phase transitions, enabling large scale MD simulations, free energy barrier calculations, and kinetic Monte Carlo simulations. 

In this study, we focus on phase transitions in MoTe$_2$ triggered by mechanical strain, as it can be conveniently applied in classical simulations based on MLIP. Nevertheless, the identified kinetic mechanisms, including both diffusive and diffusionless processes, are likely applicable to phase transitions driven by other types of external stimuli. The remainder of this paper is organized to reflect the sequence of our investigation. We begin by training a high-fidelity MLIP, followed by identifying key kinetic events associated with phase nucleation and growth using MLIP-based MD simulations, which are conducted at elevated temperature and strain levels to enhance the likelihood of observing rare events within the limited MD timescales. Next, we isolate the identified events to compute their corresponding free energy barriers and to resolve the transition pathways that are not accessible in MD simulations.  Using the extracted energy barriers, we subsequently carry out kinetic Monte Carlo (KMC) simulations to explore phase growth dynamics over extended timescales. In parallel, mean field kinetic analysis is conducted to interpret the KMC results. Finally, we summarize the identified mechanisms and discuss the implications for phase engineering in 2D TMD devices.

\section*{Results}

\subsection*{Machine learning interatomic potential}



To generate the DFT training data, we selected the SCAN exchange-correlation functional over the commonly used PBE functional due to its higher accuracy. SCAN has been shown to better reproduce the structural properties of MoTe$_2$\cite{buda2017characterization}, and in our study, it predicts significantly different potential energy landscapes compared to PBE under strain along the armchair and zigzag directions (Fig.~\ref{figs-pbe-scan}). Moreover, SCAN’s energy predictions align more closely with those from hybrid functionals\cite{Duerloo2014}. 

\begin{figure}[!t]
    \centering
    \includegraphics[width=1\linewidth]{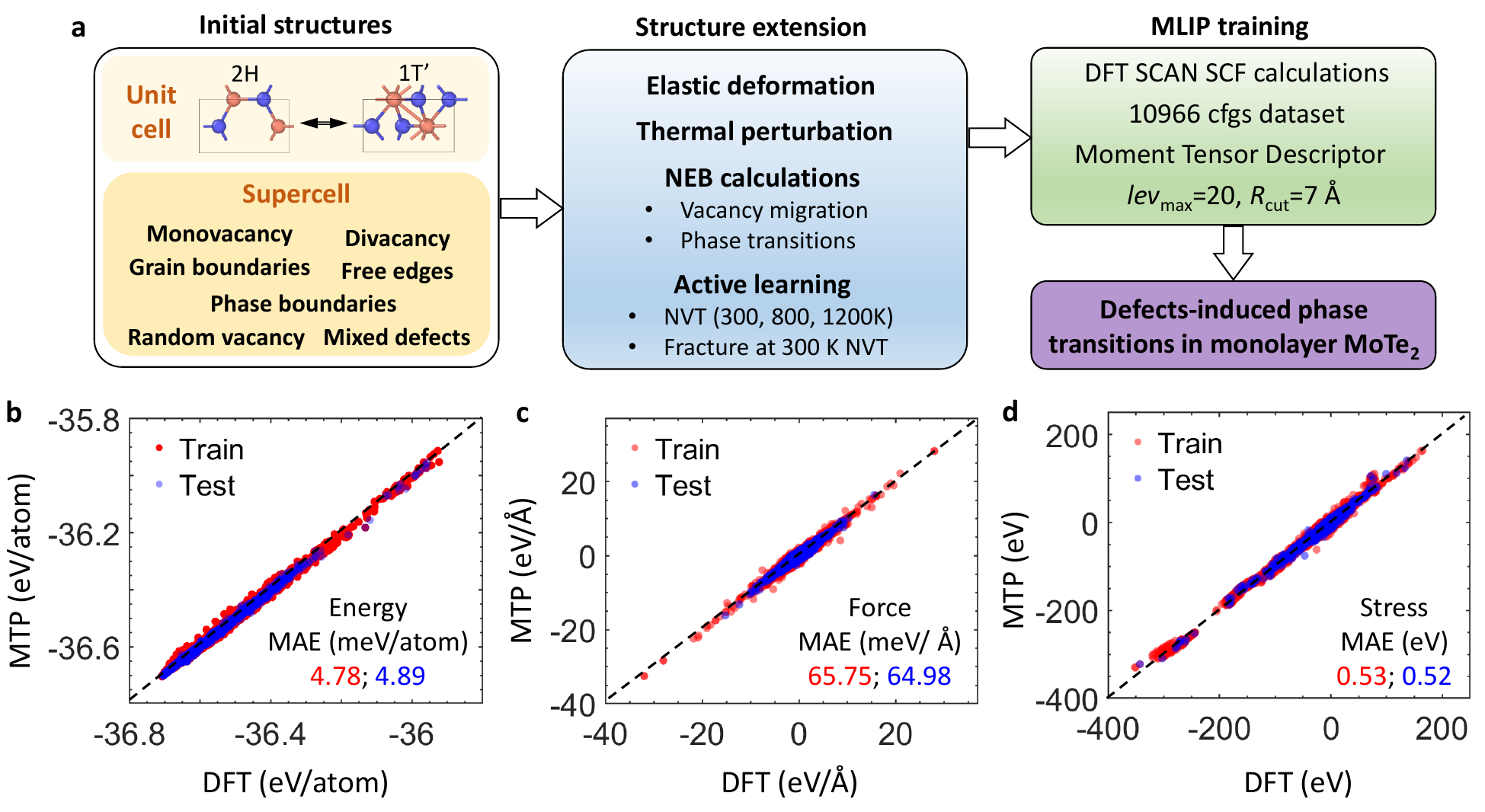}
    \caption{\textbf{Machine-learning interatomic potential development.} \textbf{a} Dataset construction and training parameters. Parity plots of the MTP prediction on \textbf{b}, energies, \textbf{c}, forces, and \textbf{d} stress for training and test data.}
    \label{fig:ML}
\end{figure}

We evaluated a total of 10,966 structures using SCAN-DFT, as summarized in Fig.~\ref{fig:ML}a. The dataset was built in two stages. First, we used domain knowledge to include a broad set of relevant configurations, such as various point and extended defects, strained structures, and transition states from nudged elastic band (NEB) calculations related to vacancy migration and the 2H$\rightarrow$1T$^\prime$ phase transition. Second, we applied on-the-fly active learning to iteratively enrich the dataset sampled from MD simulations at 300 K, 800 K, and 1200 K, including fracture simulations with armchair and zigzag cracks. The dataset was split into 90\% for training and 10\% for testing. The machine learning interatomic potential was developed using the Moment Tensor Potential (MTP) framework\cite{Novikov2021}. The optimized model achieves root-mean-square errors (RMSE) of 4.78/4.89 meV,atom$^{-1}$ for energies, 65.8/64.98 meV, $\Angstrom^{-1}$ for forces, and 0.53/0.52 eV for stresses on the training/test sets, respectively (Fig.\ref{fig:ML}b–d). We also evaluated the model’s performance on properties not explicitly included in the training or test sets, such as structural and elastic properties and phonon spectra, and found good agreement with DFT results (Fig.\ref{figs-phonon} and Table~\ref{tab:DFT-MTP comparison}).

\begin{figure}[!t]
    \centering
    \includegraphics[width=1\linewidth]{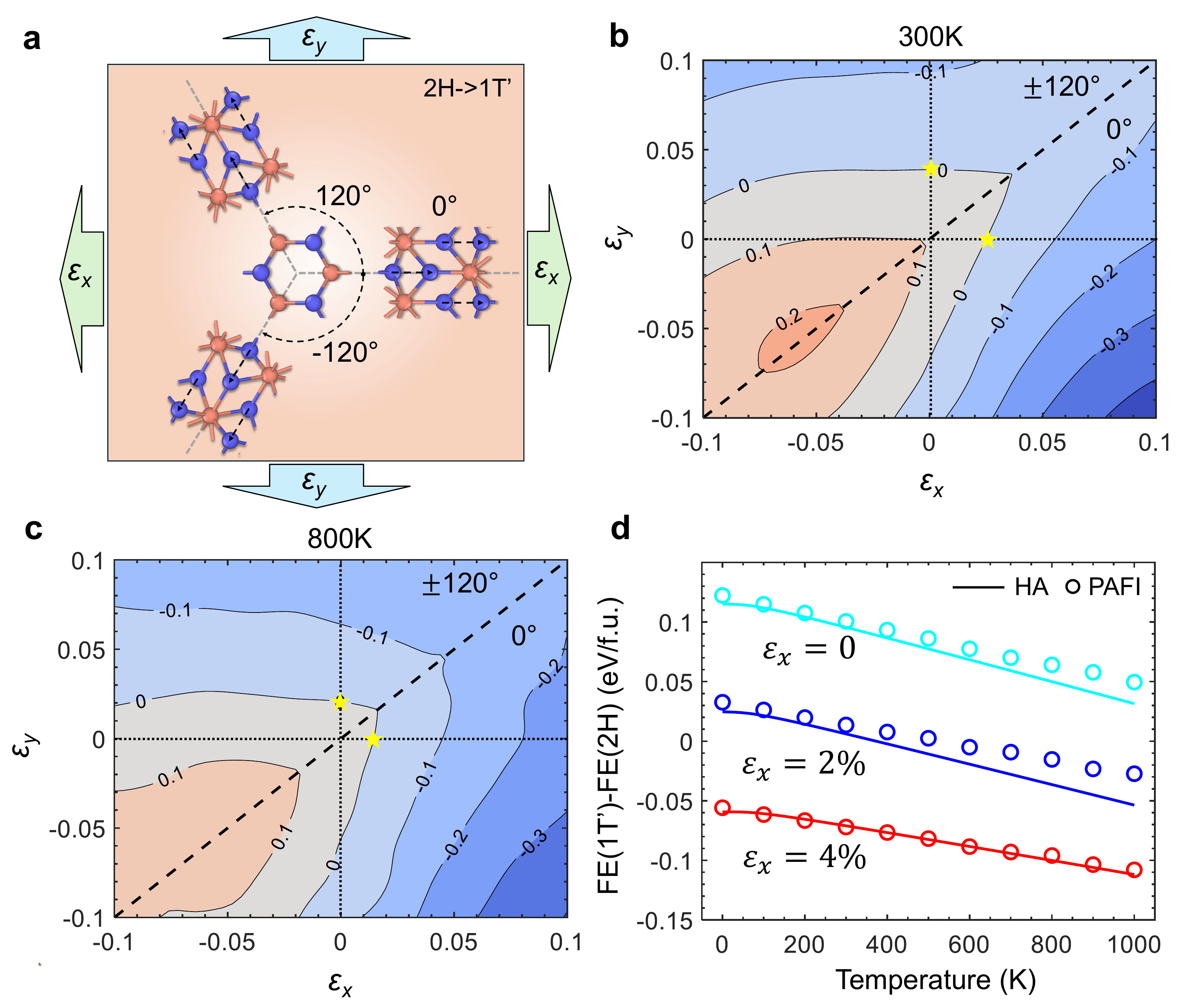}
    \caption{\textbf{Strain dependent free energy of MoTe$_2$.} \textbf{a} Three possible orientations of the 1T$^\prime$ phase with respect to the applied strain. \textbf{b-c} Free energy difference per formula between the 2H and 1T$^\prime$ phases at 300 K and 800 K. \textbf{d} Free energy difference under different armchair strains ($\varepsilon_x$) and temperatures. Results from both harmonic approximation (HA) and PAFI calculations are shown. Red and blue spheres represent Mo and Te atoms respectively.}
    \label{figs-unit-strain-temp}
\end{figure}

\subsection*{Free energy change during phase transition}

Although this study focuses on the kinetics of phase transitions, we first examine the free energy change during 2H$\rightarrow$1T$^\prime$ transition under strain using the developed MLIP, to provide a foundation for later discussions. First, we clarify a previously oversimplified conclusion regarding the effect of uniaxial strain on phase stability. Earlier studies suggested that uniaxial tension along the armchair direction promotes the transition, while tension along the zigzag direction suppresses it\cite{Duerloo2014,Ghasemi2020}. However, our results (Fig.\ref{figs-unit-strain-temp}b and \ref{figs-unit-strain-temp}c) show that tension in either direction lowers the free energy difference between the 2H and 1T$^\prime$ phases. This is due to the presence of three symmetry-equivalent 1T$^\prime$ variants, rotated by $\pm120^\circ$ with respect to the 2H lattice (Fig.\ref{figs-unit-strain-temp}a). For the off-axis orientations, the local strain components differ from the applied macroscopic strain, allowing both armchair and zigzag tension to stabilize at least one 1T$^\prime$ orientation.

Figs.~\ref{figs-unit-strain-temp}b–c show that the orientation of the 1T$^\prime$ phase can be controlled by strain: the $0^\circ$ orientation is favored when $\varepsilon_x > \varepsilon_y$, while the $\pm120^\circ$ orientations are favored when $\varepsilon_x < \varepsilon_y$. This highlights strain as a potential method for tuning phase orientation. 
Fig.~\ref{figs-unit-strain-temp}d shows that temperature lowers the energy difference but is insufficient to stabilize the 1T$^\prime$ phase; however, under 2\% strain along $x$ direction, the 1T$^\prime$ phase becomes more stable above 400 K. It is also noted that free energy values from the harmonic approximation are in close agreement with those calculated from the Projected Average Force Integrator (PAFI) method \cite{PAFI}, suggesting negligible anharmonic effects. Additional details on strain analysis and free energy calculations are provided in Supplementary Note 1.

\subsection*{Key mechanisms revealed by MLIP-MD}

Prior DFT studies on the kinetics of phase transition in MoTe$_2$ often consider concerted transformation within a unit cell. Such simplifications, however, fail to capture the complex, defect-mediated transition process. To address this, large-scale MD simulations based on MLIP are conducted on MoTe$_2$ with 5\% vacancies, subject to 4\% tensile strain along the armchair direction at 600 K. Notably, earlier DFT studies have shown that 1T$^\prime$ phase becomes thermodynamically favorable when the Te vacancy density exceeds 3\% \cite{Cho2015}. The applied 5\% vacancy level, combined with elevated temperature and strain, ensures that the system can undergo a 2H-to-1T$^\prime$ transition within the timescales accessible to MD.

The MD simulations reveal a sequence of defect-assisted processes during transition, where Te vacancy plays an important role. In particular, three key mechanisms are identified: (i) coalescence of monovacancies into divacancies, (ii) divacancy migration, and (iii) formation and growth of triangular 1T$^\prime$ island facilitated by surrounding vacancies. These key mechanisms, labeled with MD timestamps, are illustrated in Fig.~\ref{figs-md-random-vac}. In the following sections, we isolate and analyze the kinetics of each mechanism, including the corresponding free energy barriers and transition pathways, which are not directly accessible from MD simulations.

\begin{figure}[!ht]
    \centering
    \includegraphics[width=0.99\linewidth]{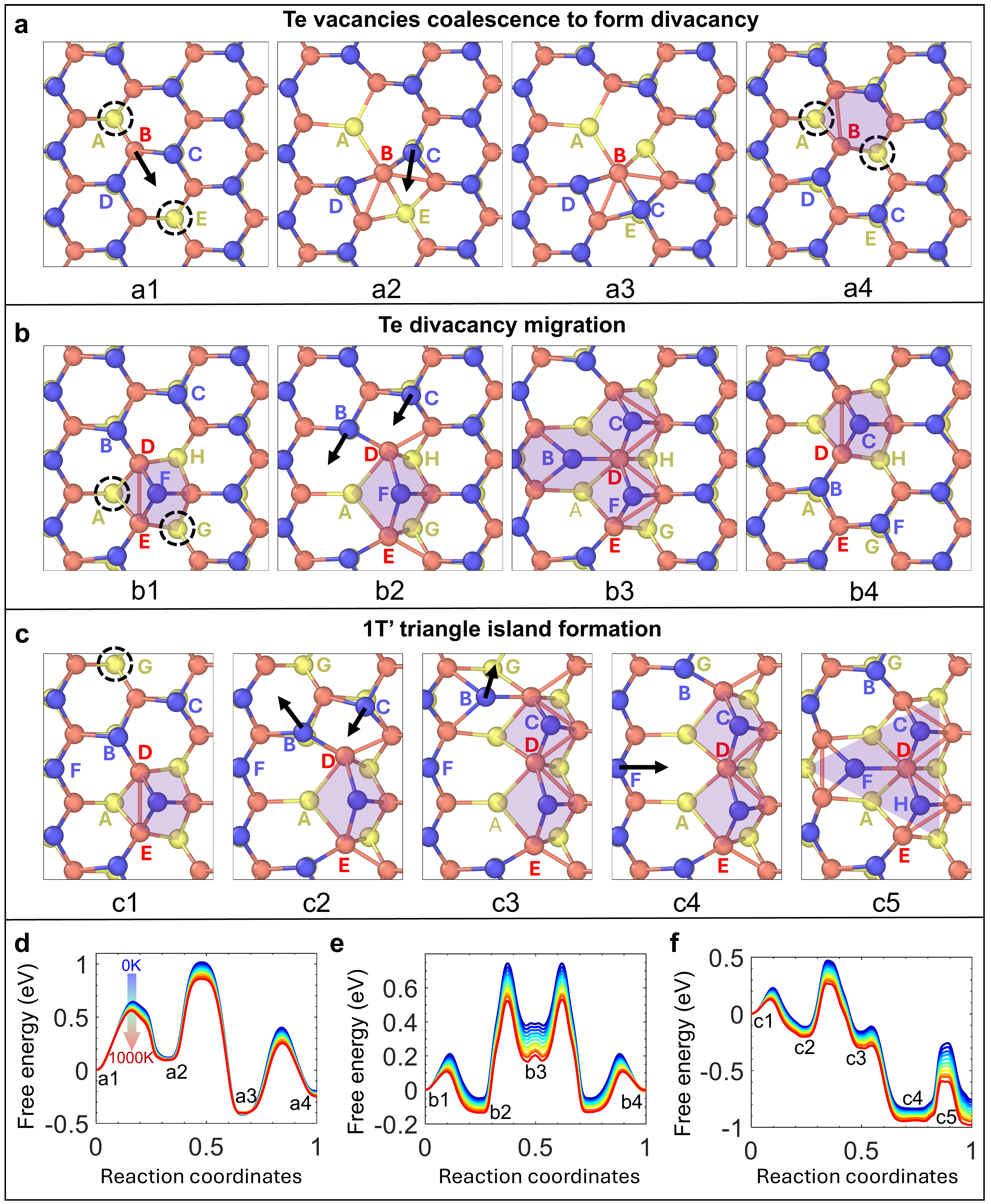}
        \vspace{0pt}
    \begingroup
     \renewcommand{\baselinestretch}{0.8}\selectfont
    \caption{\textbf{Atomistic pathway of three key transition events associated with 1T$^\prime$ phase nucleation.} \textbf{a} Coalescence of monovacancies into divacancy.  \textbf{b} Divacancy migration. \textbf{c} The formation of 1T$^\prime$ triangle island. \textbf{d-f} are the minimum free energy paths (MEPs) corresponding to these three events, respectively. Red spheres: Mo atoms; yellow and blue spheres: Te atoms in the bottom and top layers, respectively.}
    \label{fig:nucleation}
    \endgroup
    \vspace{-20pt}
\end{figure}



\subsection*{1T$^\prime$ phase nucleation from random vacancies}

The atomic pathways of the key events associated with 1T$^\prime$ nucleation, are illustrated by a series of snapshots shown in Figs.\ref{fig:nucleation}a–c. The corresponding minimum free energy paths (MEPs) are presented in Figs.\ref{fig:nucleation}d–f. The first key event, the coalescence of two monovacancies into a divacancy, is shown in Fig.~\ref{fig:nucleation}a. In the initial configuration (snapshot a1), two Te monovacancies highlighted by dashed circles, are second-nearest neighbors. The remaining Te atoms at the vacancy sites are labeled \textbf{A} and \textbf{E}. The coalescence process begins with atom \textbf{A} moving upward toward the central plane of the MoTe$_2$ monolayer, pushing Mo atom \textbf{B} along the arrow direction. This motion increases the separation between Te atoms \textbf{C} and \textbf{D}, leading to a configuration shown in snapshot a2. Subsequently, atom \textbf{C} shifts toward and sits on top of atom \textbf{E} (snapshot a3). Finally, atoms \textbf{A} and \textbf{B} relax back to their original lattice sites, completing the formation of a divacancy, as shown in snapshot a4. 
As shown in Fig.~\ref{fig:nucleation}d, the highest barrier along this pathway is 0.90~eV at 0~K, which decreases to 0.86~eV at 300~K and further to 0.76~eV at 1000~K. Notably, a 4\% tensile strain along the armchair direction applied in this calculation significantly reduces the barrier (Fig.~\ref{figs-three-strain-pafi}). Other external stimuli, such as electrostatic gating, may also lower the barrier.
Te monovacancy coalescence requires the two monovacancies to be second-nearest neighbors, which implies a locally high vacancy density. When the monovacancies are farther apart, the coalescence mechanism described above does not occur due to the high migration barrier of an isolated Te monovacancy even under large strains (Fig.\ref{figs-MV-pafi}), 
although migration of isolated Te monovacancies has been reported under electronic excitation conditions \cite{Si2019}. 

The second key event, divacancy migration, is illustrated in Fig.~\ref{fig:nucleation}b. The divacancy consists of two adjacent Te monovacancies, highlighted by the dashed circles in snapshot b1. Initially, Te atom \textbf{F}, located above Te atom \textbf{H}, moves to the center, forming a more stable divacancy core structure under strain, indicated by the shaded region. The migration of this divacancy proceeds in two steps. The first step (snapshot b2) involves structural distortion near the defect core: Te atom \textbf{A} shifts toward the central plane, pushing away two Mo atoms \textbf{D} and \textbf{E}. This is followed by Te atoms \textbf{B} and \textbf{C} moving toward the centers of adjacent hexagons, resulting in a transient 1T$^\prime$-like configuration (snapshot b3). This intermediate state corresponds to the energy minimum at the mirror-symmetric plane along the MEP (Fig.~\ref{fig:nucleation}e). The rest of transition from snapshot b3 to b4 is the reverse of the b1-b3 process. In the end, the divacancy core shown in snapshot b4 is spatially shifted relative to its initial position in snapshot b1, completing the migration step.


The third key event, the formation of a 1T$^\prime$ triangular island, is illustrated in Fig.~\ref{fig:nucleation}c. Under strain, a divacancy becomes mobile and can encounter a nearby monovacancy. As shown in snapshot c1, the divacancy is highlighted by the shaded region, while the monovacancy, circled at site \textbf{G}, is positioned as a second-nearest neighbor.  Their interaction initiates the formation of the triangular island. The process begins with Te atom \textbf{A} moving upward toward the central plane, accompanied by distortion of the divacancy core (snapshot c2). Next, Te atoms \textbf{B} and \textbf{C} in the top layer shift toward the centers of adjacent hexagons, resulting in the intermediate configuration shown in snapshot c3. Atom \textbf{B} then continues to move upward, forming configuration c4. Finally, Te atom \textbf{F} shifts laterally, completing the formation of the triangular 1T$^\prime$ island (snapshot c5). 

It is worth noting the small 1T$^\prime$ island shown in Fig.~\ref{fig:nucleation}d (snapshot c5) can revert to a vertical vacancy line upon removal of external stimuli, through sequential hopping of atoms \textbf{F}, \textbf{C}, and \textbf{H} to the left. This suggests that short Te vacancy lines and small 1T$^\prime$ islands can reversibly transform into one another. Thereby, an alternative pathway for nucleating a small 1T$^\prime$ island is via a transition from pre-existing Te vacancy lines. The corresponding energy barriers and minimum energy paths (MEPs) for this transition are shown in Fig.~\ref{figs-VL-NEB} and discussed in detail in Supplementary Note 2.

\subsection*{1T$^\prime$ phase growth via absorption of random vacancies}


In this section, we analyze the dominant process of 1T$^\prime$ phase growth following the formation of a small island. First, guided by the MD-derived intuitions, we isolate key phase growth events and compute their MEPs and activation barriers with MLIP-based NEB calculations. Then, a mean field kinetic model is proposed to describe the growth process. The theoretical predictions are compared with the results from KMC simulations, where the barriers from NEB serve as inputs. 

\begin{figure}[!t]
    \centering
    \includegraphics[width=0.99\linewidth]{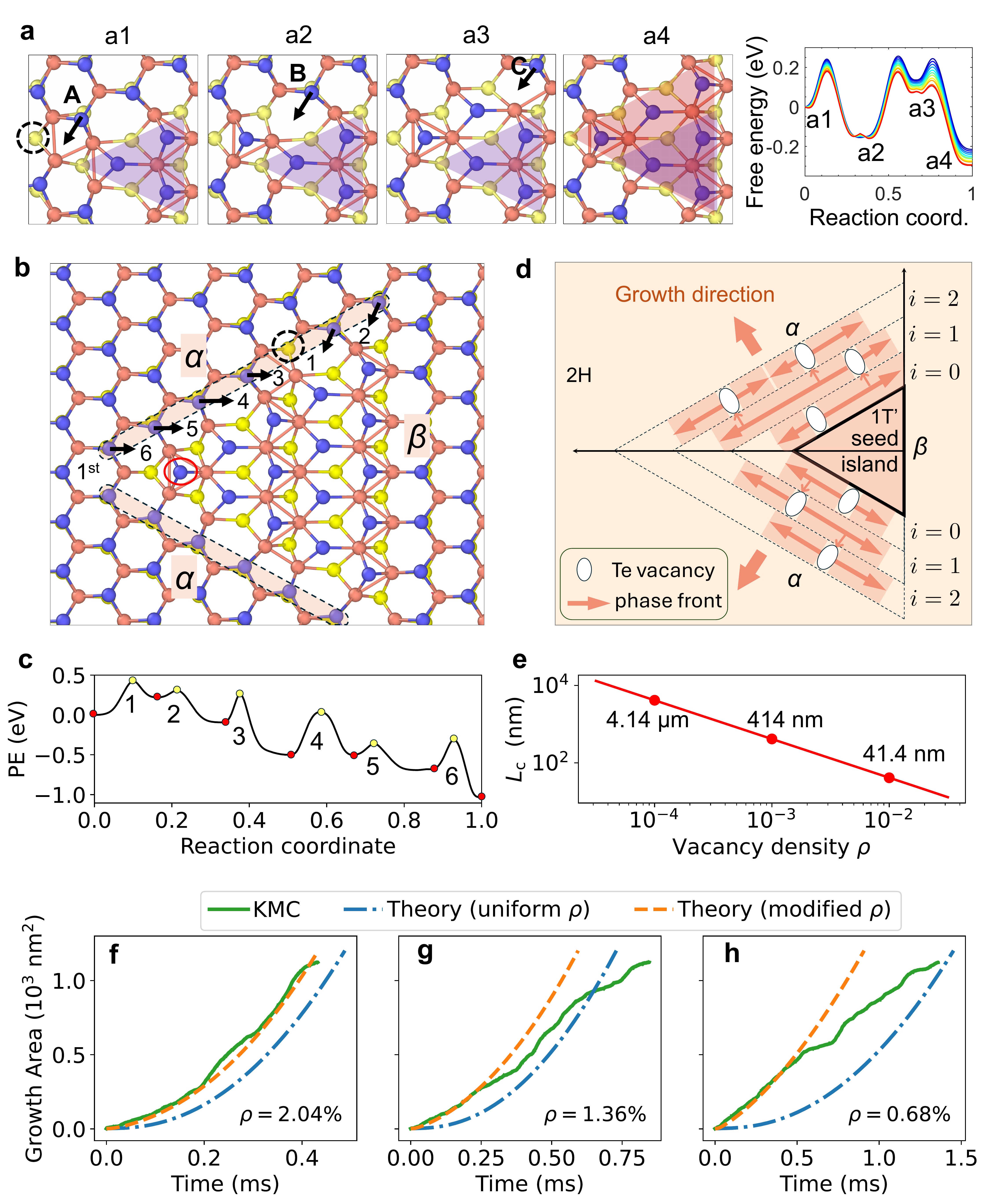}
    \vspace{-5pt}
    \caption{\textbf{1T$^\prime$ phase growth kinetics.} \textbf{a} Island growth by absorbing Te vacancies. \textbf{b} Growth on a larger system and \textbf{c} the corresponding MEP. \textbf{d} Schematics of phase front propagation, extending island row-by-row. \textbf{e} Critical island size for continuous phase growth using a threshold probability 90\%. \textbf{f-h} Evolution of growth area at different vacancy densities, comparing theory to KMC.}
    \label{fig:growth}
\end{figure}

The growth pathway is illustrated in Fig.~\ref{fig:growth}a, where a 1T$^\prime$ island is highlighted, featuring two distinct types of edge structures: a vertical edge, referred to as the $\beta$ phase boundary, and two inclined edges, referred to as the $\alpha$ phase boundaries. A Te vacancy (circled in snapshot a1) is located adjacent to the $\alpha$ edge, locally removing steric hindrance and allowing the nearby Te atom labeled \textbf{A} to hop toward the center of the hexagon. This initial hopping is followed sequentially by Te atoms \textbf{B} and \textbf{C}, eventually propagating the phase boundary forward (snapshot a2-a4). The MEP associated with this process reveals a series of low energy barriers, with a maximum value of approximately 0.4 eV, indicating that the transformation can occur readily under ambient conditions with 4\% applied armchair strain. 
To verify that the mechanism persists when the island is larger, we performed NEB calculations for a larger configuration (Fig.~\ref{fig:growth}b).  The resulting MEP (Fig.~\ref{fig:growth}c) follows the same site-by-site sequence and exhibits a comparable barrier profile.

Continued growth of the 1T$^\prime$ island proceeds through this vacancy–assisted mechanism. As illustrated in Fig.~\ref{fig:growth}d, a Te vacancy in the first row ($i=0$) outside the seed island nucleates two 1T$^\prime$ phase fronts propagating in opposite direction along the $\alpha$ edge. When either of the phase fronts reaches a Te vacancy in the next row ($i=1$), it triggers a new pair of phase fronts there; the process repeats for $i = 2,3,\dots$, expanding the island row-by-row. Next, we propose a kinetic model to quantify this growth process.

The forward and backward activation barriers corresponding to a single Te atom hopping are represented by $E_f$ and $E_b$, respectively. Then, the net 2H $\rightarrow$ 1T$^\prime$ hopping rate, $v_n$, is approximated by a difference of  Arrhenius rates: 
\begin{equation}\label{eq:rate}
    v_{\mathrm n}  = v_0\left[\exp \left(-\frac{E_f}{k_B T}\right)-\exp \left(-\frac{E_b}{k_B T}\right)\right]
\end{equation}
where $v_0$ is the attempt frequency, $k_B$ is Boltzmann constant and $T$ is temperature. 
Let $F_i(t)$ represents the number of live phase fronts in row $i$ at time $t$. Each moving front transforms the 2H lattice by Te hopping at the rate $v_\mathrm{n}$. In unit time, each front advances $v_n$ Te sites, so $F_{i-1}$ fronts in row $i-1$ sweeps across $v_n F_{i-1}$ sites. Assuming Te vacancies are independently and uniformly random distributed with density $\rho$, then a proportion $\rho$ of those sites are vacant. On average, the advancing fronts encounter $\rho v_n F_{i-1}$ vacancies per unit time. Each vacancy nucleates two counter-propagating phase fronts in row $i$. Therefore, the birth rate of the phase fronts in row $i$ can be calculated by $2\rho v_{\mathrm n}F_{i-1}$.

On the other hand, the phase fronts die when they collide or reach the end of a row. To estimate the average distance that phase fronts travel before they annihilate, we assume all $F_i$ phase fronts are evenly spaced along a row of $N_i$ Te sites, then the two end fronts each travels $N_i/(F_i+1)$ sites, while each of the remaining $F_i-2$ interior fronts travels $N_i/[2(F_i+1)]$ sites. The mean travel distance (in terms of the number of Te sites) per phase front is therefore
\begin{equation}
    \ell_\text{avg} = \left[\frac{2}{F_i}+\frac{F_i-2}{2 F_i}\right] \frac{N_i}{F_i+1}.
    \label{eq:l_avg}
\end{equation}
On average, a live phase front advances one site every $1/v_{\mathrm n}$ seconds; its mean lifetime is then $\ell_{\mathrm{avg}}/v_{\mathrm n}$, and its death rate is given by $v_{\mathrm n}/\ell_{\mathrm{avg}}$. 
In unit time, the mean number of fronts dead is $(v_\text{n}/\ell_\text{avg}) F_i$. Consider both birth and death rates, the net rate of phase fronts generation follows
\begin{equation}
    \frac{dF_i}{dt} = 2\rho v_\text{n} \,F_{i-1} (t)  - \frac{v_\text{n}}{\ell_\text{avg}} F_i (t),
    \label{eq:F_i}
\end{equation}
which give a series of first order nonlinear ordinary differential equations that can be numerically solved row-by-row starting from $i=0$. When $i=0$, $F_{-1}(t)=0$ (the seed island is static and contains no live phase fronts) and $F_0(0)= 2 \rho N_0$. When $i>0$, $F_i(0)=0$ since no fronts have yet reached row $i$ at $t=0$.

Next, let $f_i(t)$ be the fraction of Te sites in row $i$ that have already transformed to 1T$^\prime$. The temporal evolution of $f_i$ is then governed by:
\begin{equation}
  \frac{d f_i}{d t} = \frac{v_{\mathrm n}}{N_i}\,F_i(t)\,\bigl[1-f_i(t)\bigr],
  \label{eq:f_i}
\end{equation}
because, during the interval $dt$, $F_i$ phase fronts attempt $v_{\mathrm n} F_i(t) dt$ sites on average; multiplying by $(1-f_i)$ limits those attempts to the fraction of that are still 2H and therefore eligible to hop.  
This equation can be solved along with Eq.~(\ref{eq:F_i}), using $f_{-1} (t) = 1$ and $f_i (0) = 0 \; (i \geq 0)$.

Eqs.~(\ref{eq:F_i}) and (\ref{eq:f_i}) together constitute the mean field kinetic model: Eq. (\ref{eq:F_i}) propagates the phase fronts that drive the island’s growth, while Eq. (\ref{eq:f_i}) tracks how rapidly each row fills once fronts are present. Once $F_i(t)$ and $f_i(t)$ are solved, the island area is obtained by summing the contributions of all transformed sites. A single Te site approximately occupies 1T$^\prime$ area of $a_s = \frac{\sqrt{3}}{2}l_0^2$, where $l_0$ is the distance between two Te atoms (0.36 nm); row $i$ contains $N_i$ such sites and a fraction $f_i$ of them are transformed. Therefore, the total area of 1T$^\prime$ is given by $A(t) = A_0 + a_s \sum_i N_i f_i(t)$, where $A_0$ is the area of initial triangular island.

We calculated the time evaluation of 1T$^\prime$ area with both the mean field model and KMC simulations, using the same parameters detailed in Methods (KMC section). As shown in Fig.~\ref{fig:growth}f-h, the theory predicts that the area growth rate increases over time, because more vacancies become available as island expands. In the early stage, the theory (which assumes a uniform vacancy distribution) predicts a lower growth rate than KMC. Although the expected number of vacancies along the edges of a small island is below one at low vacancy densities, the growth is not halted in the theoretical predictions, because the mean field approximation represents an ensemble average over many such small islands. While most islands lack vacancies, a small fraction contains at least one, allowing growth to proceed in those cases. The predicted growth rate thus reflects an averaged value across the ensemble. This results in an initial time lag in growth. 
By contrast, in the KMC simulations, we ensure at least one vacancy per row to enable continuous growth. As a result, the vacancies are not strictly uniformly distributed, leading to immediate and faster growth of the small island during the early stage. The temporal evolution of 1T$^\prime$ island morphology from KMC simulations at different vacancy densities is shown in Fig.~\ref{figs-KMC-initial}.

As the island grows larger at later times, the theory predicts a higher growth rate than KMC. The reason is that the KMC consumes at least one forced vacancy in every rows; the remaining vacancy reservoir is therefore smaller than that expected from a uniform random distribution with the same global density. The difference in area growth rate is most pronounced at the lowest density ($\rho = 0.68\%$), where KMC enforces exactly one vacancy per row throughout the simulation. The discrepancy diminishes at $\rho=2.04\%$ because the vacancy placement in the KMC closely approximates a uniform random distribution once the island grows larger.

To make the mean field model emulate the KMC rule (at least one vacancy per row), we introduce a modified vacancy density
\begin{equation}
    \rho_\text{mod} (i) = \frac{\rho}{1-(1-\rho)^{N_i}}
\end{equation}
where the normalization factor $P_i = 1-(1-\rho)^{N_i}$ represents the probability that at least one vacancy exists along the edges of row $i$ with $N_i$ Te sites. This modification conditions the original probability $\rho$ on $P_i$, so that the modified density reflects the statistical requirement that at least one vacancy must be present before growth can proceed. When $\rho N_i \ll 1$, $P_i \simeq \rho N_i$, so $\rho_\text{mod}$ approaches to $1/N_i$, corresponding to one vacancy per row. When $\rho N_i \gg 1$, $P_i$ approaches to 1 and $\rho_\text{mod} \simeq \rho$. Thus, $\rho_\text{mod}$ smoothly interpolates between ``one vacancy per row'' condition used in KMC and the uniform bulk density $\rho$. Replacing $\rho$ in Eq.~(\ref{eq:F_i}) with $\rho_\text{mod}$ and re-solving the equations yields the results represented by dashed lines in Fig.~\ref{fig:growth}f–h. At early times, the growth rate predicted with $\rho_\text{mod}$ closely matches the KMC results across all three densities. As time progresses, the growth rate gradually approaches that predicted by the theory with uniform density. 

While the mean field theory predicts continues growth in an ensemble-averaged sense, an individual small island at low vacancy density may have a completely vacancy-free edge, causing growth to halt until a divacancy diffuses to the edge or a vacancy is externally introduced. The probability $P_i$ describes the likelihood of continued phase growth at row $i$. Because this probability approaches to 1 rapidly as the island grows, we may set a practical  threshold probability (e.g., 90\%) to mark the onset of automatic growth, we can solve for the corresponding edge length. This defines a critical length $L_c$, beyond which the island is statistically expected to grow without delay. When $\rho<<1$, this critical length approximately follows the inverse scaling law $L_c \propto 1/\rho$, as illustrated in Fig.~\ref{fig:growth}e.

\begin{figure}[!ht]
    \centering
    \includegraphics[width=1\linewidth]{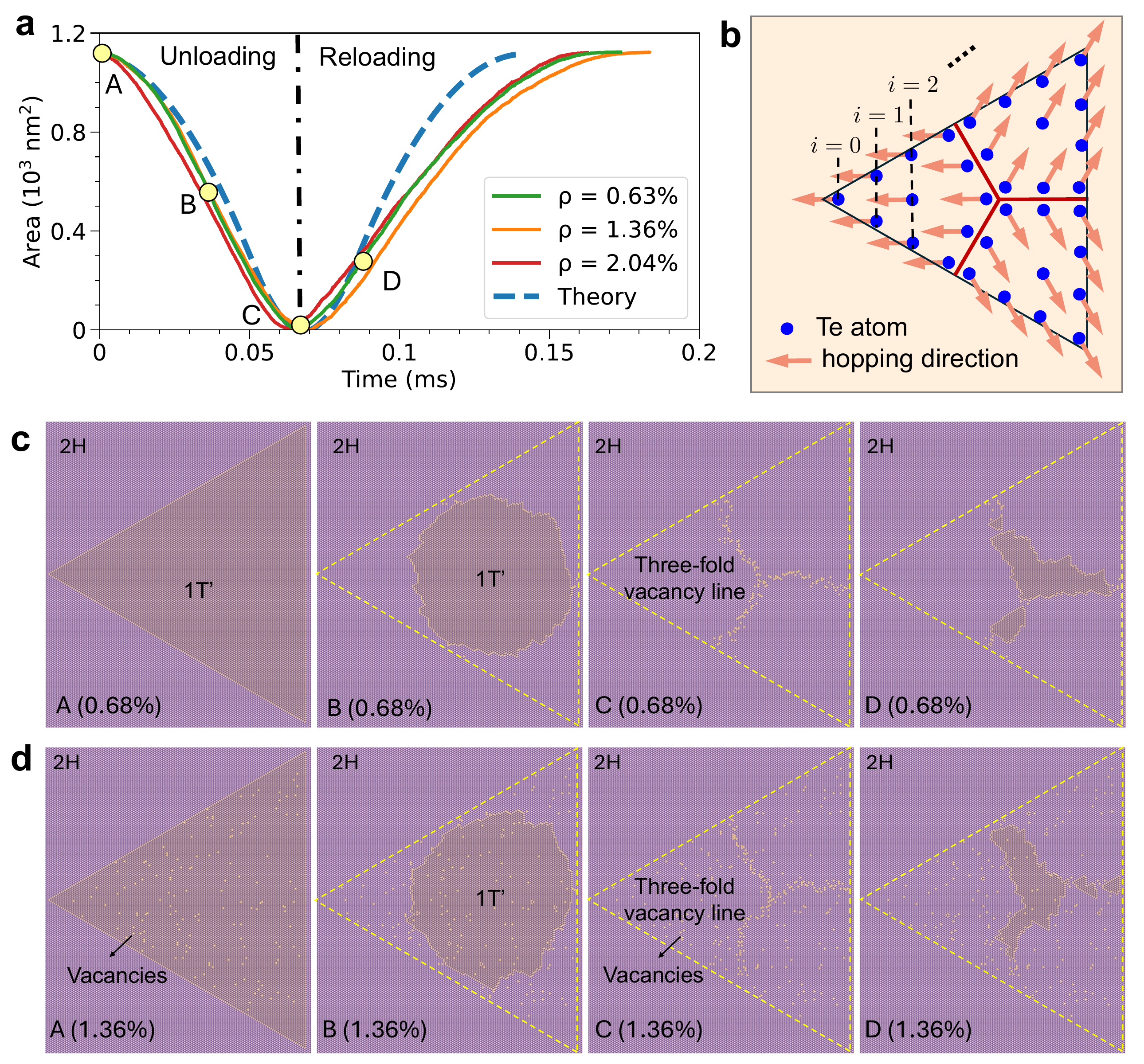}
    \caption{\textbf{Reversible phase transitions during unloading and reloading processes.} \textbf{a} Evolution of area of 1T$^\prime$ phases, comparing KMC and theory. \textbf{b} Schematic illustration of the reverse transition mechanism. \textbf{c-d} Phase morphologies at different stages from KMC.}
    \label{fig:reverse_transition}
\end{figure}

\subsection*{Diffusionless and rapid reversible phase transition}

In this section, we discuss the reverse phase transition from 1T$^\prime$ to 2H upon unloading of external stimuli (strain in our case), followed by the subsequent 2H to 1T$^\prime$ transformation upon reloading. KMC simulations are compared with a mean field model to elucidate the atomistic mechanisms.

In the KMC simulations, the initial 1T$^\prime$ island comes from the configurations generated in forward transition discussed above. Fig.~\ref{fig:reverse_transition}a shows the temporal evolution of the transformed area during unloading, and the corresponding KMC snapshots are shown in Figs.\ref{fig:reverse_transition}c and d. There are four observations. (1) The reverse transition follows a sequence of atomic rearrangements opposite to that of the forward transition. However, unlike the forward process, which is gated by the probabilistic presence of Te vacancies, the reverse transition proceeds rapidly without such vacancy constraints. (2) The reverse transition initiates from the corners of the 1T$^\prime$ island and propagates toward the center, forming an intermediate, circular morphology (snapshot B in Fig.\ref{fig:reverse_transition}c–d). (3) At the end of reverse transition, the Te vacancies that facilitated the forward transition remain in the 2H phase. Most are positioned along three spoke-like lines forming a 3-fold pattern (snapshot C), while the rest are randomly distributed. (4) The reverse growth rate is weakly dependent on the vacancy density in the given dilute regime. The vacancies in KMC are not uniformly distributed, since they are inherited from the preceding forward step.

Fig.~\ref{fig:reverse_transition}b illustrates the reverse transition process. Since the Te atoms at the three corners of the triangular island are free from steric hindrance, they hop first, followed by the atoms in adjacent rows in a row-by-row fashion. Let the Te atoms in row $i$ be the children, and those in row $i-1$ the parents.  Hopping of an interior child atom requires two parent atoms to hop first in order to relieve steric hindrance, while an edge child (two edge atoms per row) requires only one. Therefore, the atoms along the edges transform faster than interior ones, leading to the circular shaped pattern as seen in KMC. This behavior can be described by the following mean field kinetic equation:
\begin{equation}\label{eq:reverse_growth}
    \frac{df_i}{dt} = v_n \left( \frac{2}{N_i}f_{i-1} + \frac{N_i-2}{N_i} f_{i-1}^2 \right) (1-f_i),
\end{equation}
where $v_n$ is the net 1T$^{\prime}$$\rightarrow$2H hop rate of one site,  $f_i (t)$ is the fraction of Te sites hopped at row $i$ by time $t$, and $N_i$ is the total number of sites in row $i$. The terms $f_i-1$ and $f_{i-1}^2$ represent the conditional probability associated with one-parent and two-parent dependencies, respectively. The reverse transition propagates simultaneously from all three corners and converges at the spokes in the center, leaving Te vacancies along the spoke lines. The theoretical predictions from numerical integration of Eq.(\ref{eq:reverse_growth}) agree closely with KMC results, as shown in Fig.\ref{fig:reverse_transition}a. The reverse growth rate first increases while the transitions run along two $\alpha$ edges, then decreases as they converge along the spokes. In the reverse transition, many Te atoms can hop in parallel once their parent sites have moved, in contrast to the forward transition, where hopping proceeds one-by-one along the phase fronts. As a result, the reverse transition is one order of magnitude faster than forward one and does not rely on vacancies to facilitate the process.

Upon reloading, the system undergoes a 2H$\rightarrow$1T$^\prime$ transition, following a reverted pathway of previous 1T$^\prime$$\rightarrow$2H transition. The growth initiates from the center and expands toward the corners (reversing the hopping directions in Fig.\ref{fig:reverse_transition}b).  The kinetics also follow Eq.(\ref{eq:reverse_growth}), with one difference: Te atoms along the edge now require two hopped parent atoms, while those at the spokes need only one, owing to the presence of nearby Te vacancies. As shown in Fig.~\ref{fig:reverse_transition}a, the theory overestimates the growth rate compared to KMC results, mainly due to the theoretical assumption that all vacancies lie exactly along the spoke lines. In reality, as seen in KMC, the vacancies are distributed within a narrow band surrounding the spokes.
This reduces the effective number of sites that require only one parent, thereby slowing the growth. In addition, it results in asymmetric growth among the three wedges (separated by the spoke lines) with mutual interference, in contrast to the symmetric and independent wedge growth assumed in the theoretical model.

\subsection*{Alternative 1T$^\prime$ phase growth pathways}

\begin{figure}[!b]
    \centering
    \includegraphics[width=1\linewidth]{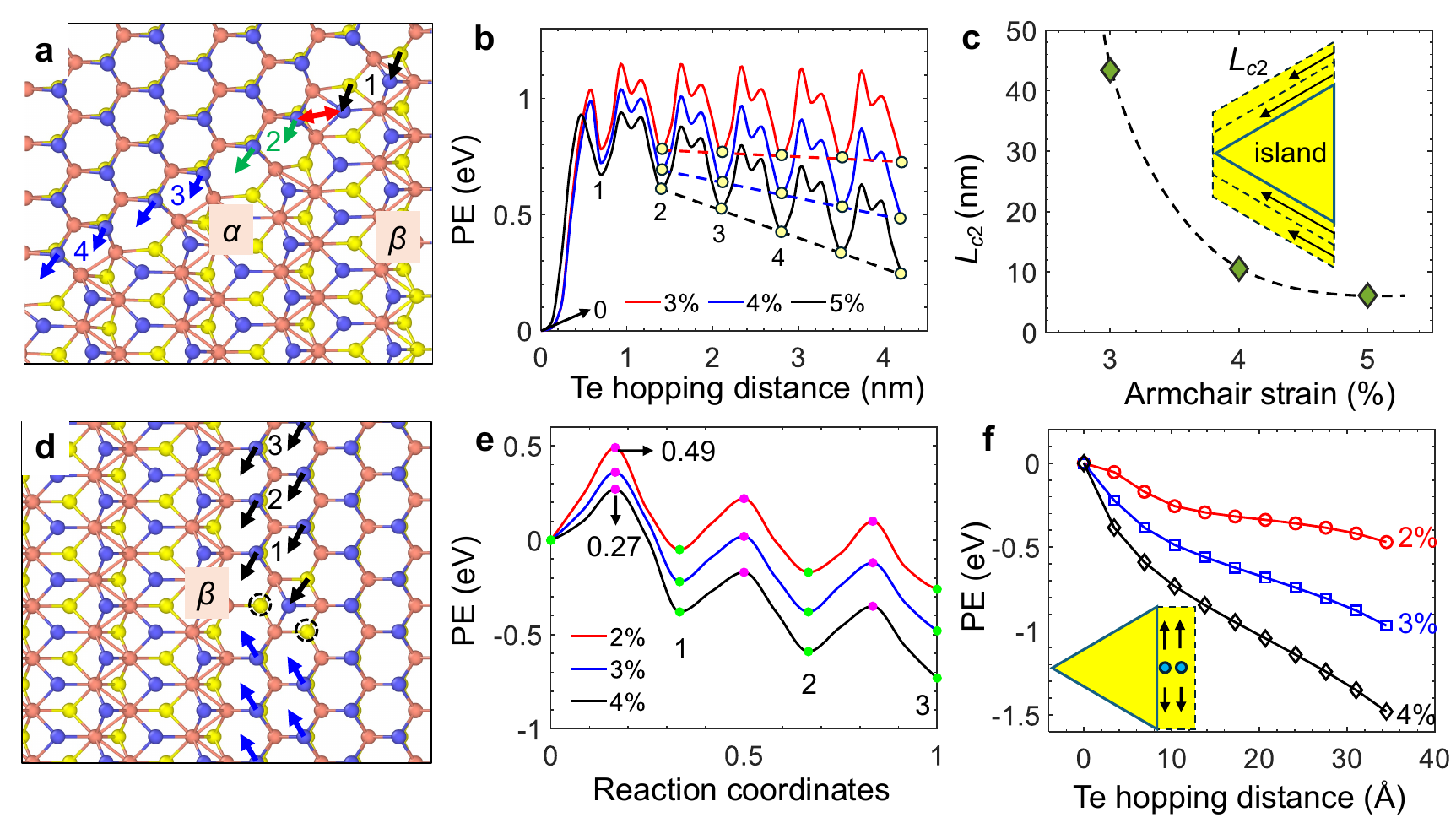}
    \caption{\textbf{Alternative 1T$^\prime$ phase growth mechanisms.} \textbf{a} Atomistic mechanisms of 1T$^\prime$ phase growth without vacancies, \textbf{b} the corresponding minimum energy paths as Te atoms hopping along the $\alpha$ edge, and \textbf{c} the critical size of 1T$^\prime$ triangle island. \textbf{d} Atomistic mechanisms of 1T$^\prime$ triangle island growth with a divacancy near the $\beta$ phase boundary, \textbf{e} the corresponding minimum energy paths of first three hopping events, and \textbf{f} potential energy change versus Te atoms hopping distance along the $\beta$ edge.}
    \label{fig:secondary_growth}
\end{figure}

In the previous sections, we discussed the primary 1T$^\prime$ growth mode, in which the phase expands as a triangular island along two inclined $\alpha$ phase boundaries. This mechanism requires at least one Te vacancy at two advancing fronts. For sufficiently large islands, this condition is highly likely to be met even at dilute vacancy densities. Therefore, we consider it the dominant growth pathway. In this section, we introduce two alternative growth modes, which were also observed in our MD and tfMC simulations (Figs. \ref{figs-md-no-defect} and \ref{figs-tfmc-no-defect}). Their pathways impose more stringent requirements and are therefore considered secondary growth mechanisms.

The first alternative pathway is the vacancy-free edge growth (Fig.\ref{fig:secondary_growth}a–c). This mechanism does not rely on vacancies but must overcome a higher initial energy barrier. As illustrated in Fig.\ref{fig:secondary_growth}a, growth initiates with the concerted displacement of two Te atoms (arrows labeled as 1), resulting in a sharp energy increase from state 0 to state 1 (Fig.~\ref{fig:secondary_growth}b). This is accompanied by the formation of a perfect phase boundary that imposes strong Te–Te steric repulsion (highlighted by the red arrow). However, once this is established, subsequent transformations simply translate the boundary laterally with a series of low energy barriers. Our calculations in Fig.~\ref{figs-pafi-no-defect} show that the free energy barrier for the initial move drops to 0.90 eV at 800 K. This growth mode was observed in MD simulations at the same temperature (Fig.~\ref{figs-md-no-defect}), indicating that although the process is kinetically sluggish, it can be thermally activated under external stimuli.

The MEPs in Fig.~\ref{fig:secondary_growth}b show that the system’s energy decreases as Te atoms progressively hop along the $\alpha$ edge. However, if the edge length of the initial triangular 1T$^\prime$ island is too short, the energy of the extended 1T$^\prime$ phase remains higher than the initial state, making the transition thermodynamically unfavorable. There exists a critical 1T$^\prime$ edge length beyond which growth becomes energetically favorable. We estimate this critical size from the MEPs in Fig.~\ref{fig:secondary_growth}b (extending the dash lines) and plot it in Fig.~\ref{fig:secondary_growth}c. Under 4\% armchair strain, continuous growth is achievable for 1T$^\prime$ islands with edge lengths on the order of 10 nm. Once the initial island exceeds this critical size, the vacancy-free growth mechanism becomes possible and leads to the formation of 1T$^\prime$ band structure (shown as the inset in Fig.~\ref{fig:secondary_growth}c), which is different from the triangular island morphology seen in the vacancy-assisted growth.

The second alternative phase growth mode is $\beta$ phase boundary growth assisted by divacancies. As shown in Fig.~\ref{fig:secondary_growth}d, this transition occurs along the $\beta$ boundary in the presence of an adjacent Te divacancy, where two Te atoms collectively migrate into neighboring hollow sites, forming a 1T$^\prime$ unit cell. Once triggered, the sequential Te atoms hopping propagates upward and downward along the vertical $\beta$ edge. Notably, a single Te vacancy on $\beta$ edge can only initiate a partial hopping sequence and therefore cannot complete a full 1T$^\prime$ unit cell. The associated energy barriers are low even at small strains (Fig.~\ref{fig:secondary_growth}e), and the system’s potential energy decreases as the front advances (Fig.~\ref{fig:secondary_growth}f), first vertically along the $\beta$ boundary and then laterally in the armchair direction with additional divacancies, leading to a growth morphology (inset in Fig.~\ref{fig:secondary_growth}f) different from previous two. 
Because the probability of finding a Te divacancy adjacent to a $\beta$ boundary is relatively low, this growth mechanism relies on the migration of divacancies. As such, it is a diffusion-driven process.

\sisetup{
  scientific-notation = true,
  table-format=1.2e2 
}

\section*{Summary and Discussion}

\begin{figure}[!b]
    \centering
    \includegraphics[width=1\linewidth]{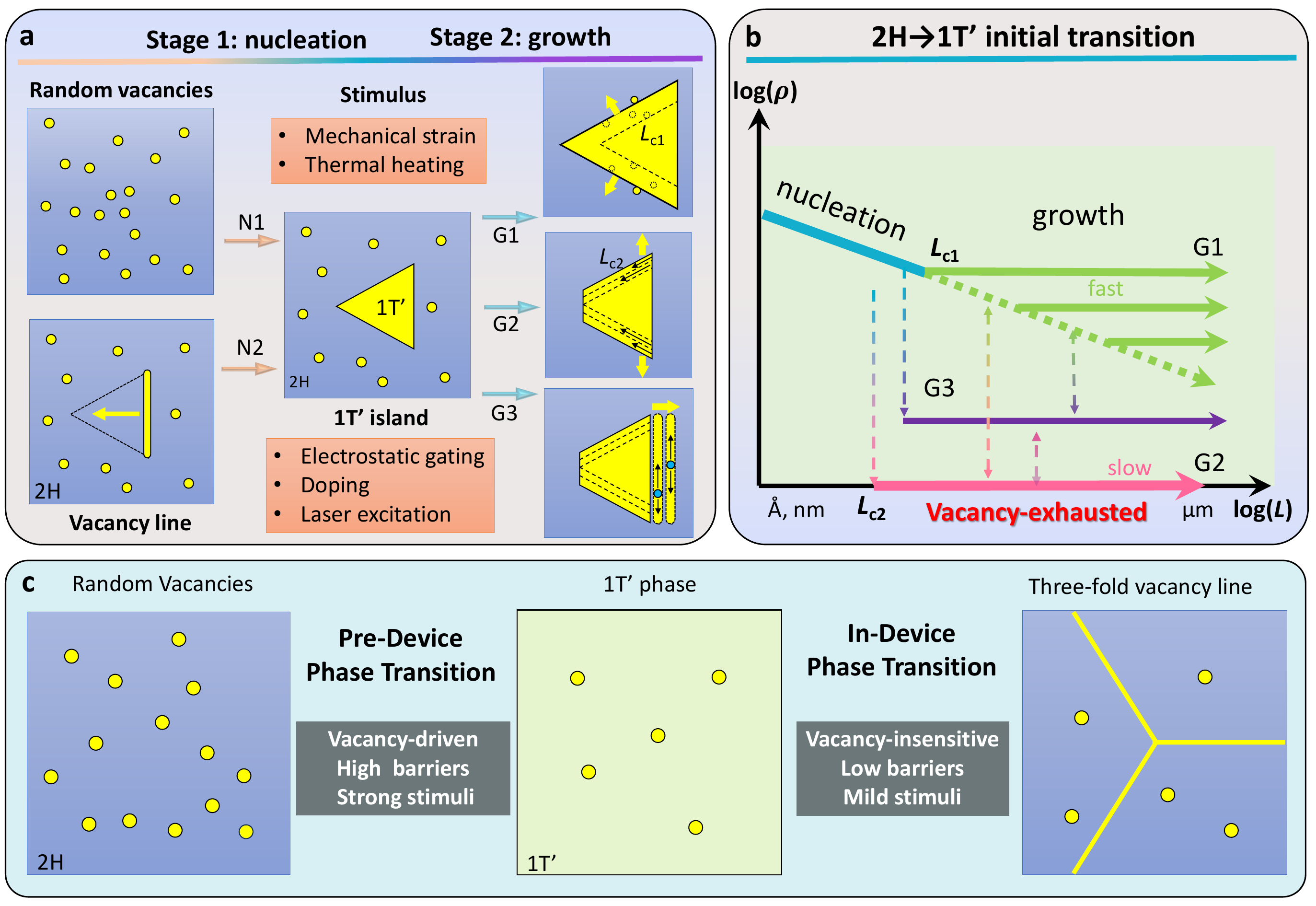}
    \caption{\textbf{Summary of the kinetic mechanisms in 2H-1T$^\prime$ phase transition.} \textbf{a} Schematic of the mechanisms of the initial transition, and \textbf{b} their dependence on vacancy concentration. \textbf{c} Two-stage phase engineering approach.}
    \label{fig:summary}
\end{figure}

The key findings of this study are summarized in Fig.\ref{fig:summary}. The initial 2H $\rightarrow$ 1T$^\prime$ transition can be divided into two stages: nucleation and growth (Fig.\ref{fig:summary}a). During nucleation, small 1T$^\prime$ triangular islands form either through the interactions of Te vacancies (N1) or direct transition from pre-existing vacancy lines (N2). After nucleation, the 1T$^\prime$ islands expand through three distinct growth mechanisms, labeled G1, G2, and G3, each leading to a different growth morphology. 
The activation barriers and estimated timescales of key kinetic events, summarized in Table~\ref{tab:time_scale}, indicate that growth is significantly more kinetically favorable than nucleation.

The role of vacancies is illustrated in Fig.~\ref{fig:summary}b. Nucleation requires a relatively high local concentration of Te vacancies to increase the likelihood of forming small 1T$^\prime$ islands through vacancy interactions. Once the island reaches a sufficient size, the probability of encountering at least one vacancy along each of the two $\alpha$ phase boundaries increases significantly, even at dilute vacancy densities, enabling continued growth via the G1 mode.  However, if vacancies are exhausted, further growth relies on the migration of divacancies. By contrast, the G2 mode allows vacancy-free growth once the island exceeds a critical size, though this pathway involves a higher activation barrier. Finally, the G3 mode supports growth by absorbing divacancies at the $\beta$ boundary, making the process explicitly dependent on divacancy migration.

It is important to highlight that the kinetics of the initial 2H$\rightarrow$1T$^\prime$ transition are significantly different from those of the reverse transition upon unloading and the subsequent transitions under cyclic loading. The initial transition may involve both diffusive and diffusionless processes, whereas the later reversible transitions proceed purely through diffusionless mechanisms. 
Motivated by this difference, we propose a two-stage phase engineering strategy, illustrated in Fig.~\ref{fig:summary}c. The pre-device stage corresponds to the initial transition, which requires a locally high vacancy concentration to trigger the transformation from the 2H phase. Due to the relatively high activation barriers, strong external stimuli, such as large strains, elevated temperatures, or large gating voltages, are necessary. Once this stage is complete, a stable 2H/1T$^\prime$ heterostructure is formed, enabling the next in-device stage. In this stage, removing the external stimulus triggers a rapid reverse transition (1T$^\prime$$\rightarrow$2H), during which the phase front retreats from the three corners toward the center of the island, leaving behind three spoke-like vacancy lines. Subsequent reversible transitions, switching between these vacancy lines and the 2H/1T$^\prime$ heterostructure, proceed rapidly through continuous domain growth, without requiring new nucleation events or additional vacancies.

Finally, we highlight an important implication of our study for experimental efforts. While many experiments in recent years have demonstrated successful phase transitions in 2D TMD materials under various external stimuli, a gap remains between laboratory discoveries and practical phase engineering applications. Reproducibility may be challenging, particularly when researchers attempt to replicate published results under similar conditions. We cautiously hypothesize, although without evidence, that some previous experiments may have unknowingly operated in the in-device stage, where phase transitions can be achieved using only mild stimuli. In such cases, the more demanding pre-device transition might have already occurred during material fabrication or through unintended exposure to strong stimuli. As a result, attempts to reproduce these transitions in 2H crystals that have not undergone the pre-device transition may fail when only mild stimuli are applied. This underscores the importance of distinguishing the kinetic characteristics of the pre-device and in-device stages when designing experiments and interpreting results.

\section*{Methods}

\subsection*{DFT calculations}
Our \textit{ab initio} calculations are carried out with the Vienna \textit{Ab Initio} Simulation Package (VASP) using the Strongly Constrained and Appropriately Normed (SCAN) meta‑GGA functional\cite{Sun2015}. SCAN reproduces the energy difference between the 2H and 1T$^\prime$ allotropes with an accuracy comparable to HSE06 and superior to PBE (Fig.~\ref{figs-pbe-scan}).  We enforce an electronic self‑consistency tolerance of \(10^{-6}\,\text{eV}\).  Structural optimisations converge when the residual forces drop below $(10^{-3}\,\text{eV\,}\Angstrom^{-1}$); for CI‑NEB images we relax until the forces fall below \(2\times10^{-2}\,\text{eV\,}\Angstrom^{-1}\).  All calculations employ a kinetic‑energy cutoff of 400\,eV.  \textbf{k}‑point meshes are generated with \textsc{VasPkit}\cite{VASP-KIT} according to the Monkhorst–Pack scheme, at a uniform density of \(2\pi\times0.03\,\Angstrom^{-1}\) across the entire dataset.

\subsection*{MLIP framework}

The moment tensor-based MLIP\cite{Novikov2021} is used to construct contracted rotationally invariant local environment descriptors for each atom in the system and builds a polynomial regressed correlation between the potential energy surface (PES) and these descriptors. The descriptors, named moment tensors, are devised as follows:
\begin{equation} \label{eq11}
M_{\mu,\nu}(R) = \sum_j \left[
        \underbrace{
            f_{\mu}(|R_{ij}|, z_i, z_j)
        }_{\text{radial}}
        \cdot 
        \underbrace{
            \overbrace{
                R_{ij} \otimes \cdots \otimes R_{ij}
            }^{\nu \text{ times}}
        }_{\text{angular}}
\right]
\end{equation}
where the function $f_{\mu}$ are the radial distributions of the local atomic environment around atom $i$, specified to the neighboring atom $j$. The term $R_{ij} \otimes \cdots \otimes R_{ij}$ are tensors of rank $\nu$, encoding the angular information about the local environment. There are two key parameters that determine the accuracy and computational cost of the trained MLIP: the cutoff radius ($R_{\text{cut}}$) and the maximum level ($lev_\text{{max}}$). In this work, we choose $R_{\text{cut}} = 7\, \Angstrom$ and $lev_\text{{max}} = 20$. The energy, force and stress data are assigned weights of 1, 0.01, and 0.001. During active learning, we monitor the extrapolation grade of MTP to gauge model reliability. Configurations whose grade exceeds 2.1 are automatically harvested and added to the training set for on‑the‑fly retraining, whereas a threshold of 10 aborts the MD run to prevent sampling configurations that lie far outside the potential’s domain of validity.

\subsection*{MD and tfMC simulations}
Classical MD simulations are performed with \textsc{LAMMPS}\cite{Thompson2022} in the canonical (NVT) ensemble at 600\,K.  
Uniaxial tension is applied along both the armchair and zigzag directions to a simulation cell of \(\sim25\times25\;\text{nm}^2\) containing \(\sim16\,000\) atoms, using a time step of 1.0\,fs. The simulation results of one case are presented in Fig. \ref{figs-md-random-vac}. To accelerate sampling of rare phase‑transition events, we also employ the time-stamped force-bias Monte Carlo (tfMC) simulations implemented in LAMMPS \cite{Thompson2022}.  
The key parameter in tfMC is the displacement amplitude \(\Delta\), which sets the magnitude of the random atomic kicks; here we choose \(\Delta=0.25\) to aggressively perturb all atoms and expedite the dynamics of vacancy‑rich 2H MoTe$_2$. The temperature is set as 800 K. This approach enables access to the atomistic mechanisms of the 2H-1T$^\prime$ transition that would be inaccessible on conventional MD timescales. The simulation results of one case are presented in Fig. \ref{figs-tfmc-random-vac}.

\subsection*{Free energy barrier calculations}
Free energy minimum energy paths are computed with the Projected Average Force Integrator (PAFI)\cite{PAFI}.  We discretise the NEB pathway into 31 hyperplanes and sample each at nine temperatures spanning 0–1000 K in 100 K increments.  On every hyperplane the system is first equilibrated for 1000 MD steps under overdamped Langevin dynamics (friction coefficient 0.001), followed by 1000 production steps from which the mean constraint force is collected.  PAFI launches one independent worker per hyperplane–temperature pair; each worker runs on a single CPU, giving 384 concurrent sampling threads in the present calculations. The ensemble of averaged forces is subsequently integrated to yield the free‑energy barrier.

\subsection*{Kinetic Monte Carlo (KMC) simulations}\label{sec:kmc}
We performed lattice‐based KMC simulations to study phase growth in monolayer MoTe$_2$, neglecting explicit atomic and lattice relaxations. For each possible Te hop $i$, the rate is $v_i = v_0 \cdot \text{exp}({-\left( E_i / (k_B \cdot T) \right)})$, where $k_0$, $k_\text{B}$ and $T$ denote the attempt frequency ($1.5 \times 10^{13} \, \text{s}^{-1}$), Boltzmann constant ($8.61733326 \times 10^{-5}$ eV/K), and simulation temperature, respectively. $E_i$ represents the hopping barrier along the jump path $i$, which includes both forward and backward transitions.
For 2H$\rightarrow$1T$^\prime$ growth, the forward hopping barrier varies depending on the Te site’s position (Fig.~\ref{fig:growth}a). We adopt a representative value of $E_f = 0.4$ eV, corresponding to the highest barrier among the sequential transitions. The backward barrier is offset by the free energy difference between phases at 4\% strain at room temperature, $\Delta E_\text{cell}=0.02$ eV, giving $E_b=E_f + \Delta E_\text{cell} = 0.42$ ev. The same values for $E_f$, $E_b$ and $v_0$ are used in the mean field model for comparison.

In KMC, the total jump rate is the sum of all individual rates, $R = \sum_{i=1}^{s} v_i$, where $s$ denotes the number of available hops. A uniform random number $u$ within the interval (0,1] selects event $p$ that meets the following condition: $\sum_{i=1}^{p-1} \frac{v_i}{R} \leq u \leq \sum_{i=1}^{p} \frac{v_i}{R}$. The time increment associated with that event is estimated by $\Delta t = -\ln(r) / R$, with a random number $0 < r < 1$.

\clearpage 

%
\bibliography{ref} 
\bibliographystyle{sciencemag}

%
%
%
%
%
%

\newpage
\section*{Acknowledgments}
W.G. gratefully acknowledges financial support of this work by the National Science Foundation through Grant no.
CMMI-2308163 and CMMI-2305529. The authors acknowledge the Texas Advanced Computing Center (TACC) at the University of Texas at Austin and Texas A\&M High Performance Research Computing for providing HPC resources that have contributed to the research results reported within this paper. 

\paragraph*{Author contributions:}
F.S. contributed to methodology, software development, investigation, original draft writing, and review and editing of the manuscript. D.O., R.N., P.D., and A.G. contributed to investigation and review and editing of the manuscript. W.G. contributed to conceptualization, methodology, investigation, original draft writing, review and editing of the manuscript, supervision, and funding acquisition.
\paragraph*{Competing interests:}
There are no competing interests to declare.
\paragraph*{Data and materials availability:}
All source codes, DFT dataset, and MLIP are available at the GitHub repository: \url{https://github.com/ufsf/ML-MoTe2}.

\subsection*{Supplementary materials}
Supplementary Notes 1-3\\
Figs. S1 to S12

\newpage


\renewcommand{\thefigure}{S\arabic{figure}}
\renewcommand{\thetable}{S\arabic{table}}
\renewcommand{\theequation}{S\arabic{equation}}
\renewcommand{\thepage}{S\arabic{page}}
\setcounter{figure}{0}
\setcounter{table}{0}
\setcounter{equation}{0}
\setcounter{page}{1} 


\begin{center}
\section*{Supplementary Materials for\\ \scititle}

Fei Shuang,
Daniel Ocampo,
Reza Namakian,\\
Arman Ghasemi,
Poulumi Dey,
Wei Gao$^\ast$\\
\small$^\ast$Corresponding author. Email: wei.gao@tamu.edu\\
\end{center}

\subsubsection*{This PDF file includes:}
Supplementary Notes 1-3\\
Figures S1 to S12

\newpage

\subsubsection*{Supplementary Note 1. Phase transitions without defects}

Using our newly developed machine‑learning interatomic potential (MLIP), we examine the intrinsic 2H-1T$^\prime$ transformation in a single MoTe$_2$ unit cell (Fig.~\ref{figs-unit-strain-temp}).  Earlier first‑principles work proposed that uniaxial tension along the armchair axis promotes the 2H-1T$^\prime$ conversion, whereas tension along the zigzag axis suppresses it\cite{Duerloo2014}.  By contrast, our results in Fig.~\ref{figs-unit-strain-temp}b and \ref{figs-unit-strain-temp}c show that both loading directions lower the energy difference. The apparent discrepancy arises because the 1T$^\prime$ lattice admits three symmetry‑equivalent variants rotated by \(0^{\circ}\) and \(\pm120^{\circ}\) with respect to the 2H frame.  For the two off‑axis variants the resolved strains along the local \(a\) (armchair) and \(b\) (zigzag) axes differ from the macroscopic strain, enabling either armchair or zigzag tension to stabilize at least one 1T$^\prime$ orientation.

According to the strain transformation in a 2D plane, the strain experienced by the transitioned 1T$^\prime$ phase can be expressed as:

\begin{equation}
\left\{
\begin{aligned}
\varepsilon_{x'} &= \frac{\varepsilon_x + \varepsilon_y}{2} + \frac{\varepsilon_x - \varepsilon_y}{2} \cos 2\theta + \varepsilon_{xy} \sin 2\theta, \\
\varepsilon_{y'} &= \frac{\varepsilon_x + \varepsilon_y}{2} - \frac{\varepsilon_x - \varepsilon_y}{2} \cos 2\theta - \varepsilon_{xy} \sin 2\theta, \\
\varepsilon_{x'y'} &= -\frac{\varepsilon_x - \varepsilon_y}{2} \sin 2\theta + \varepsilon_{xy} \cos 2\theta,
\end{aligned}
\right.
\end{equation}
where $\theta$ is the rotation angle, and $\varepsilon_{x}$ and $\varepsilon_{y}$ are the applied macroscopic strains along the $x$ and $y$ axes, respectively. In our analysis we set the in‑plane shear component to zero, \(\varepsilon_{xy}=0\), because shear deformation can trigger wrinkling instabilities in freestanding monolayers. The strains $\varepsilon_{x'}$, $\varepsilon_{y'}$, and $\varepsilon_{x'y'}$ correspond to the strains experienced by the 1T$^\prime$ phase along its $a$ direction, $b$ direction, and in-plane shear, respectively. When $\theta = 0\degree$, the strain experienced by the transitioned 1T$^\prime$ phase matches the applied macroscopic strain. However, when $\theta = 120\degree$ or $-120\degree$, the strains become:

\begin{equation}
\left\{
\begin{aligned}
\varepsilon_{x'} &= \frac{\varepsilon_x}{4} + \frac{3}{4}\varepsilon_y, \\
\varepsilon_{y'} &= \frac{3}{4}\varepsilon_x + \frac{\varepsilon_y}{4}, \\
\varepsilon_{x'y'} &= \pm\frac{\sqrt{3}}{4}(\varepsilon_x - \varepsilon_y).
\end{aligned}
\right.
\end{equation}
This indicates that macroscopic strain along the $y$ direction ($\varepsilon_y$) can also induce significant tensile strain on the $120\degree$-oriented 1T$^\prime$ phase ($\varepsilon_{x'}$), facilitating the phase transition. Additionally, non-zero shear strain may arise in the $120\degree$-oriented 1T$^\prime$ phase if $\varepsilon_x$ and $\varepsilon_y$ are unequal.

To analyze the effects of strain and temperature, we calculate the free energy differences between the 2H and 1T$^\prime$ phases for three 1T$^\prime$ orientations (i.e., $F[\text{1T$^\prime$}(0\degree)] - F[\text{2H}(0\degree)]$, $F[\text{1T$^\prime$}(120\degree)] - F[\text{2H}(0\degree)]$, and $F[\text{1T$^\prime$}(-120\degree)] - F[\text{2H}(0\degree)]$) using the harmonic approximation over a wide strain range at temperatures of 300 K and 800 K. The free energy difference diagrams, along with the thermodynamically preferred 1T$^\prime$ orientations, are plotted in Figs.~\ref{figs-unit-strain-temp}b and \ref{figs-unit-strain-temp}c. These results demonstrate that both armchair and zigzag strains effectively reduce the free energy difference between the 2H and 1T$^\prime$ phases. Notably, the thermodynamically favored orientation of the transitioned 1T$^\prime$ phase depends on the applied strain. Increasing $\varepsilon_x$ and decreasing $\varepsilon_y$ stabilize the $0\degree$-oriented 1T$^\prime$ phase, while decreasing $\varepsilon_x$ and increasing $\varepsilon_y$ stabilize the $120\degree$-oriented 1T$^\prime$ phase. Specifically, the $0\degree$-1T$^\prime$ orientation is preferred when $\varepsilon_x > \varepsilon_y$, whereas the $\pm120\degree$-1T$^\prime$ orientation is favored when $\varepsilon_x < \varepsilon_y$. This suggests a potential method for controlling the orientation of the 1T$^\prime$ phase through strain engineering. Furthermore, Figs.~\ref{figs-unit-strain-temp}b and \ref{figs-unit-strain-temp}c illustrate that applying strain along the $x$ direction is more effective in facilitating the phase transition than along the $y$ direction. The critical strains for the transition are determined to be $\varepsilon_{xc} = 1.8\%$ and $\varepsilon_{yc} = 2.8\%$ at 300 K, and $\varepsilon_{xc} = 1.4\%$ and $\varepsilon_{yc} = 2\%$ at 800 K. The minimum strain required for the transition can be further reduced under stress-controlled uniaxial loading.

\begin{figure}[!ht]
    \centering
    \includegraphics[width=1\linewidth]{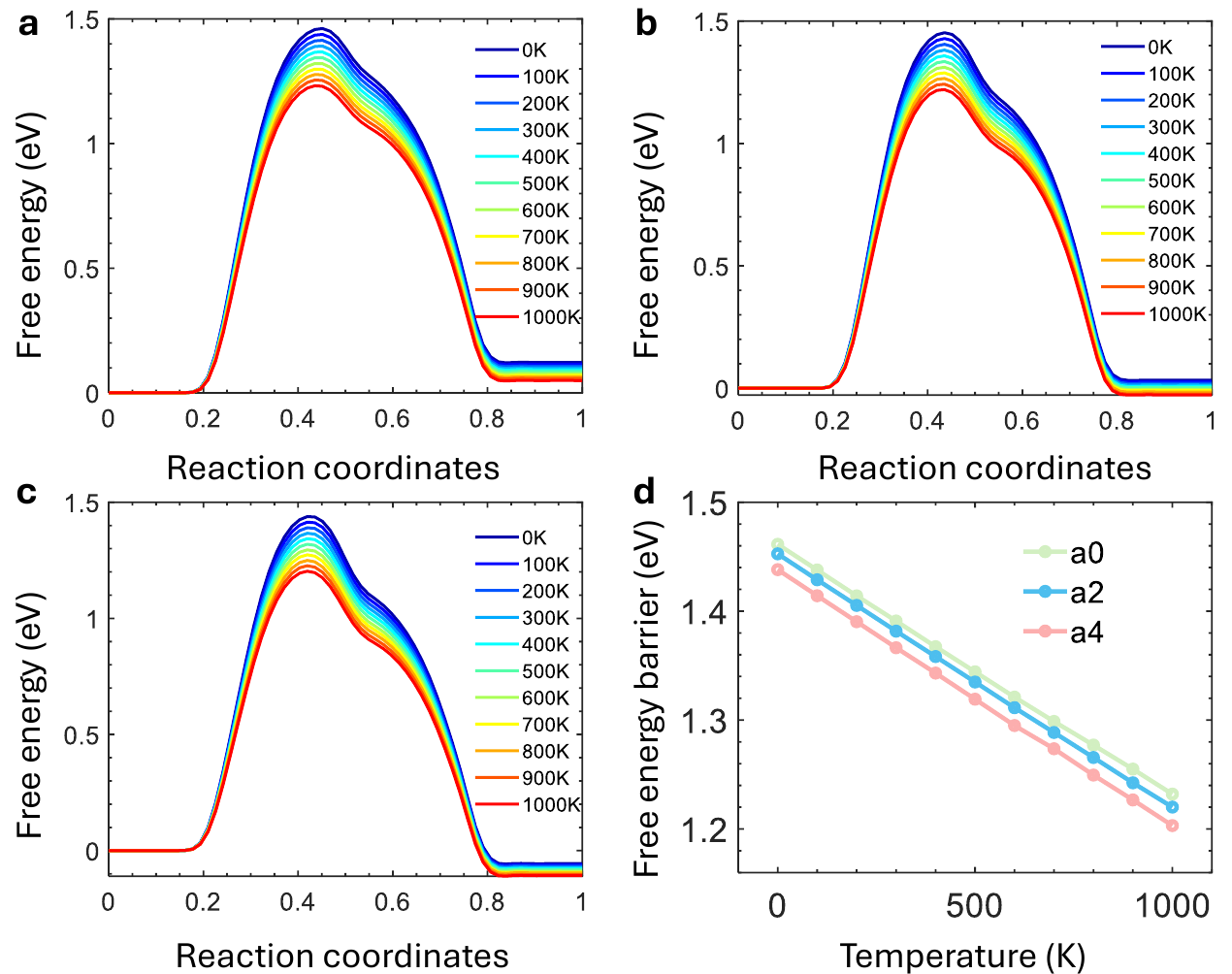}
    \caption{\textbf{Free energy calculations of phase transition from 2H to 1T$^\prime$ in the form of unit cell (concerted mechanism)}. \textbf{a}-\textbf{c} Minimum free energy path at temperatures 0-1000 K under strain conditions of 0, 2\%, and 4\% along the armchair direction. \textbf{d} Temperature-dependent free energy barriers under different strain conditions.}
    \label{figs-unit-pafi}
\end{figure}

To account for anharmonic effects in free energy calculations, we employ the PAFI method \cite{PAFI} to compute the minimum free energy path for the phase transitions of the unit cell under various strains. The free energy differences between the 2H and 1T$^\prime$ phases, plotted in Fig.~\ref{figs-unit-strain-temp}d, reveal that anharmonic effects are negligible and are suppressed at high strains of 4\%. These results indicate that the thermodynamic requirement can be easily satisfied in a unit cell. However, Fig.~\ref{figs-unit-pafi} presents the free energy barriers for the 2H$\rightarrow$$0\degree$-1T$^\prime$ transition obtained via the PAFI method. These barriers exceed 1.2 eV under extreme temperature (1000 K) for a unit cell, even at very high strains, indicating a potential for fracture induction prior to phase transition. While other stimuli, such as electrostatic gating, doping, and charging, may reduce the energy barrier to a smaller value, they cannot eliminate it entirely. This limitation prevents these methods from explaining the large-area 2H$\leftrightarrow$1T$^\prime$ phase transitions observed experimentally. Considering both experimental and simulation results, we conclude that defect-free concerted phase transitions are not feasible.

\newpage

\subsubsection*{Supplementary Note 2. Phase transitions from a vacancy line in 2H phase}

\begin{figure}[!ht]
    \centering
    \includegraphics[width=1\linewidth]{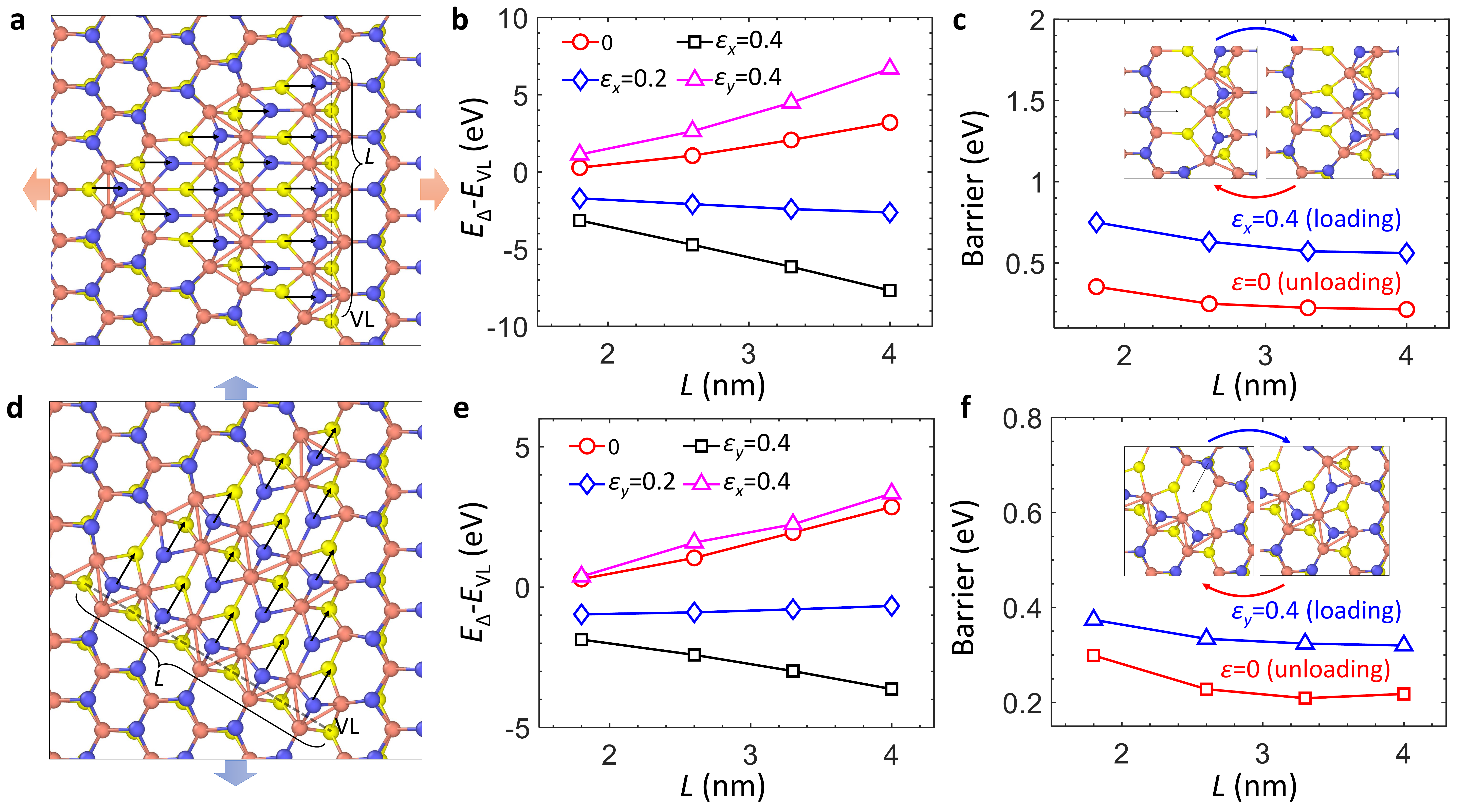}
    \caption{\textbf{Formation of the 1T$'$ phase triangle from a Te vacancy line (VL).}  
    \textbf{a, d} Atomistic mechanisms of the VL-to-1T$^\prime$ phase transition driven by armchair-oriented (a) and zigzag-oriented (d) strain.  
    \textbf{b, e} Energy difference between the vacancy line and the 1T$^\prime$ triangular island as a function of VL length and applied strain.  
    \textbf{c, f} Critical energy barriers for the phase transition corresponding to different VL lengths.}  
    
    \label{figs-VL-NEB}
\end{figure}

Another critical mechanism for the formation of a 1T$^\prime$ triangular island is the growth from a vacancy line (VL) in 2D TMDs, as evidenced by experimental observations of VL existence \cite{Lehnert2019}. The transition from a VL to a 1T$^\prime$ triangular island involves a sequential shift of Te atoms toward the original VL, as illustrated in Figs.~\ref{figs-VL-NEB}a and \ref{figs-VL-NEB}d. The energy difference \((E_{\triangle} - E_{\text{VL}})\), highlighted in Fig.~\ref{figs-VL-NEB}b, underscores the thermodynamic favorability of 1T$^\prime$ island formation at different strain levels. A minimum strain of \(\varepsilon_{x} = 2.0\%\) along the armchair direction is required to stabilize the 1T$^\prime$ island, as shown in Fig.~\ref{figs-VL-NEB}a. A VL of length \(L\) can transform into a 1T$^\prime$ triangular island with an equivalent edge length, with larger VLs leading to greater energy reductions during this process. However, applying a large strain of \(\varepsilon_{y} = 4.0\%\) along the zigzag direction is detrimental to this transition, consistent with the strain-dependent phase stability shown in Fig.~\ref{fig:nucleation}.

Fig.~\ref{figs-VL-NEB}c presents the energy barriers for a representative transition event across different VL lengths, demonstrating that longer VLs require lower barriers for individual Te atom shifts. We also analyze the barriers for the backward transition from a 1T$^\prime$ triangle to a VL during the unloading process at zero strain, which are lower than those for the forward transition. Fig.~\ref{figs-VL-NEB}e indicates that applying a zigzag strain of \(\varepsilon_{y} = 4.0\%\) facilitates the formation of a 1T$^\prime$ island with the armchair direction aligned along the \(120\degree\) orientation, whereas an armchair strain of \(\varepsilon_{x} = 4.0\%\) impedes it. Climbing Image Nudged Elastic Band (CI-NEB) calculations in Fig.~\ref{figs-VL-NEB}f reveal that the barriers for both forward and backward transitions are low and decrease with increasing VL length.

Overall, these findings suggest that the orientation of the 1T$^\prime$ island can be controlled through different combinations of mechanical strain, and the 2H-to-1T$^\prime$ phase transition is reversible by applying or removing these strains. It is important to note that the VL-to-triangular island transition not only provides a 1T$^\prime$ nucleus but also represents a potential pathway for large-area phase transitions from 2H to 1T$^\prime$, provided the VL is sufficiently long. Given that all barriers during this process are relatively low, as shown in Figs.~\ref{figs-VL-NEB}c and \ref{figs-VL-NEB}f, the transition between a VL and a large 1T$^\prime$ triangular phase is expected to be fast and controllable. We estimate the timescale for such a process to be on the order of 0.1 ms. However, a limitation of this mechanism is the requirement for a long VL, which may not always exist in experiments or practical applications without manual intervention.

\newpage

\subsubsection*{Supplementary Note 3. MD and tfMC simulation results}
\begin{figure}[!ht]
    \centering
    \includegraphics[width=1\linewidth]{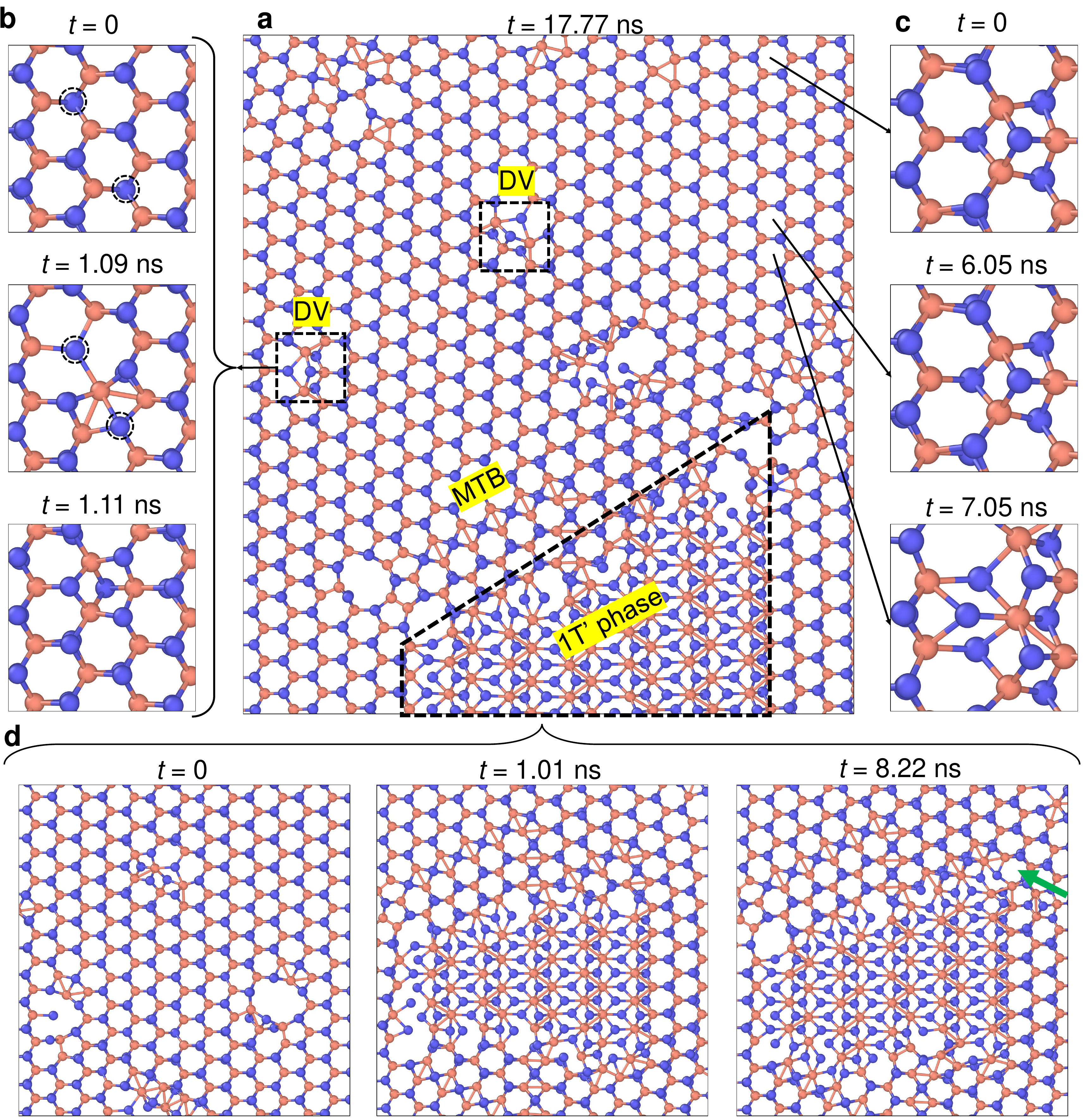}
    \caption{\textbf{MD simulation results of phase transitions in defective 2H phase}. \textbf{a} Final structure after 17.77 ns. \textbf{b} Te monovacancy coalescence. \textbf{c} Te divacancy migration. \textbf{d} 1T$^\prime$ phase nucleation from a local region with high concentration of Te vacancies.}
    \label{figs-md-random-vac}
\end{figure}

We began with large‑scale MD simulations to identify the possible atomistic mechanisms of the 2H-1T$^\prime$ transformation. All runs were performed in the canonical (NVT) ensemble at 600 K, with a 4\% tensile strain applied along the armchair direction to accelerate the process. After 17.77 ns, the defective 2H lattice evolved into the configuration shown in Fig.~\ref{figs-md-random-vac}a, revealing three critical events: (i) coalescence of Te monovacancies (MVs) into divacancies (Fig.~\ref{figs-md-random-vac}b); (ii) long‑range divacancy migration (Fig.~\ref{figs-md-random-vac}c); and (iii) growth of triangular 1T$^\prime$ domains (Fig.~\ref{figs-md-random-vac}d). These observations provide a direct atomistic picture of vacancy‑mediated phase transformation in MoTe$_2$. Furthermore, we estimate the timescales of these three events are in the order of 1.0 ns, which are significantly faster than the results in Table 1 obtained by transition state theory. This discrepancy could be due to the presence of other defects.

\newpage

\subsubsection*{Figures}
\begin{figure}[!ht]
    \centering
    \includegraphics[width=1\linewidth]{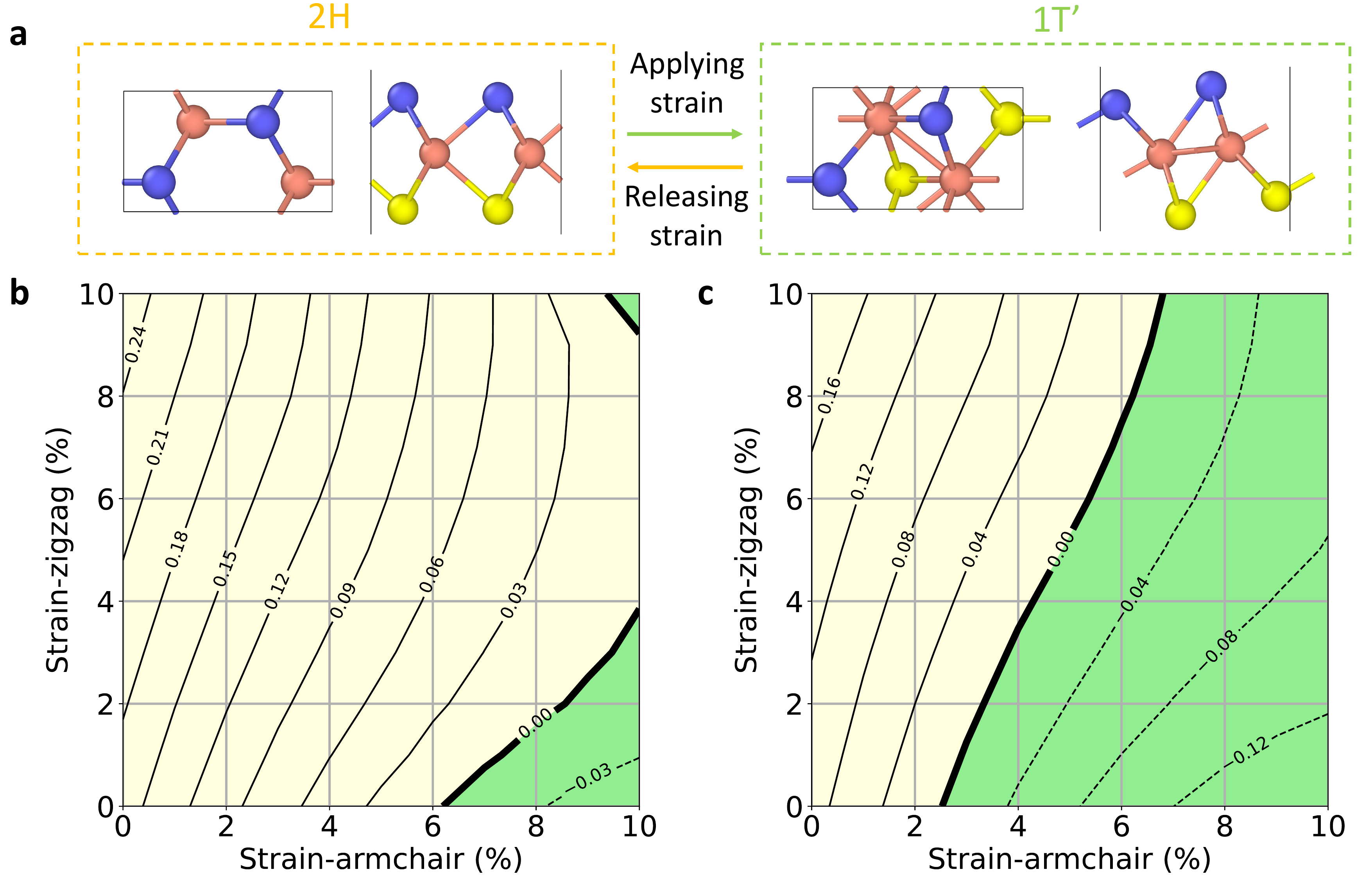}
    \caption{\textbf{Thermodynamics of phase transition between 2H and 1T$^\prime$ phase}. \textbf{a} Mechanical strain-induced phase transition in unit cell of 2H and 1T$^\prime$ phase. \textbf{b} Contour plot of energy difference between 2H and 1T$^\prime$ using PBE exchange functional. \textbf{c} Contour plot of energy different between 2H and 1T$^\prime$ using SCAN exchange functional.}
    \label{figs-pbe-scan}
\end{figure}

\newpage

\begin{figure}[!ht]
    \centering
    \includegraphics[width=1\linewidth]{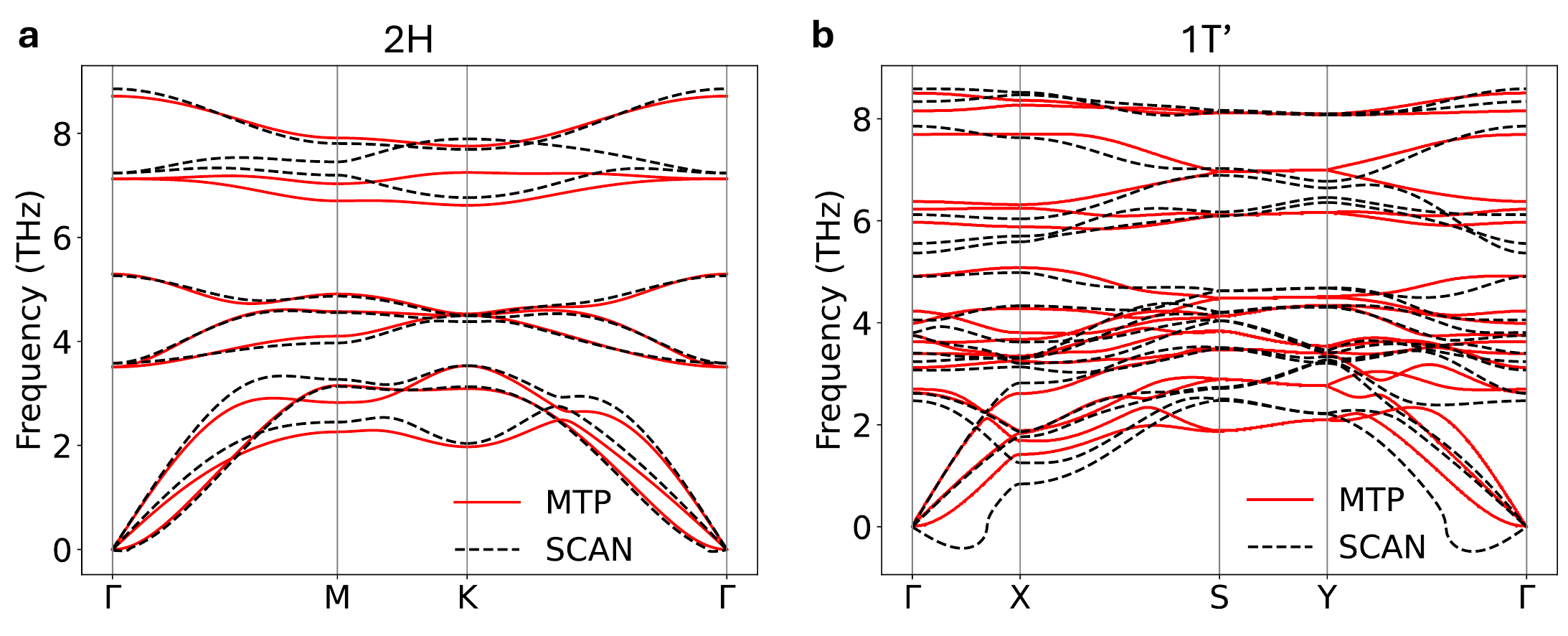}
    \caption{\textbf{Phonon spectra validation of machine-learning interatomic potential (MLIP) for 2H and 1T$^\prime$ phases.} Comparison of phonon dispersion relations calculated using density functional theory (DFT) and MTP demonstrates excellent agreement for both (a) 2H-MoTe$_2$ and (b) 1T$^\prime$-MoTe$_2$ phases, validating the accuracy of our MLIP in capturing lattice dynamics.}
    \label{figs-phonon}
\end{figure}

\vspace{30pt}

\begin{table}[htbp]
\centering
\caption{Comparison of structural and elastic properties from DFT and MTP models}
\begin{tabular}{lllll}
\toprule
\textbf{Phase} & \textbf{Property} & \textbf{SCAN-DFT} & \textbf{MTP} & \textbf{Error (\%)} \\
\midrule
\multirow{6}{*}{1T$^\prime$} 
& $a (\Angstrom)$   & 3.375    & 3.413    & 1.126 \\
& $b (\Angstrom)$   & 6.379    & 6.349    & 0.470 \\
& $t (\Angstrom)$   & 4.081  & 4.091  & 0.237 \\
& $C_{11}$ (N/m) & 117.050 & 105.275 & 10.060 \\
& $C_{12}$ (N/m) & 35.771  & 32.000  & 10.541 \\
& $C_{22}$ (N/m) & 84.962  & 98.090  & 15.451 \\
\midrule
\multirow{5}{*}{2H} 
& $a (\Angstrom)$   & 3.511    & 3.511    & 0.003 \\
& $b (\Angstrom)$   & 6.082    & 6.082    & 0.002 \\
& $t (\Angstrom)$   & 3.583    & 3.582    & 0.023 \\
& $C_{11}$ (N/m) & 91.339   & 89.945   & 1.526 \\
& $C_{12}$ (N/m) & 20.153   & 23.779   & 17.995 \\
\bottomrule
\end{tabular}
\label{tab:DFT-MTP comparison}
\end{table}

\newpage

\newpage

\begin{figure}[!ht]
    \centering
    \includegraphics[width=0.7\linewidth]{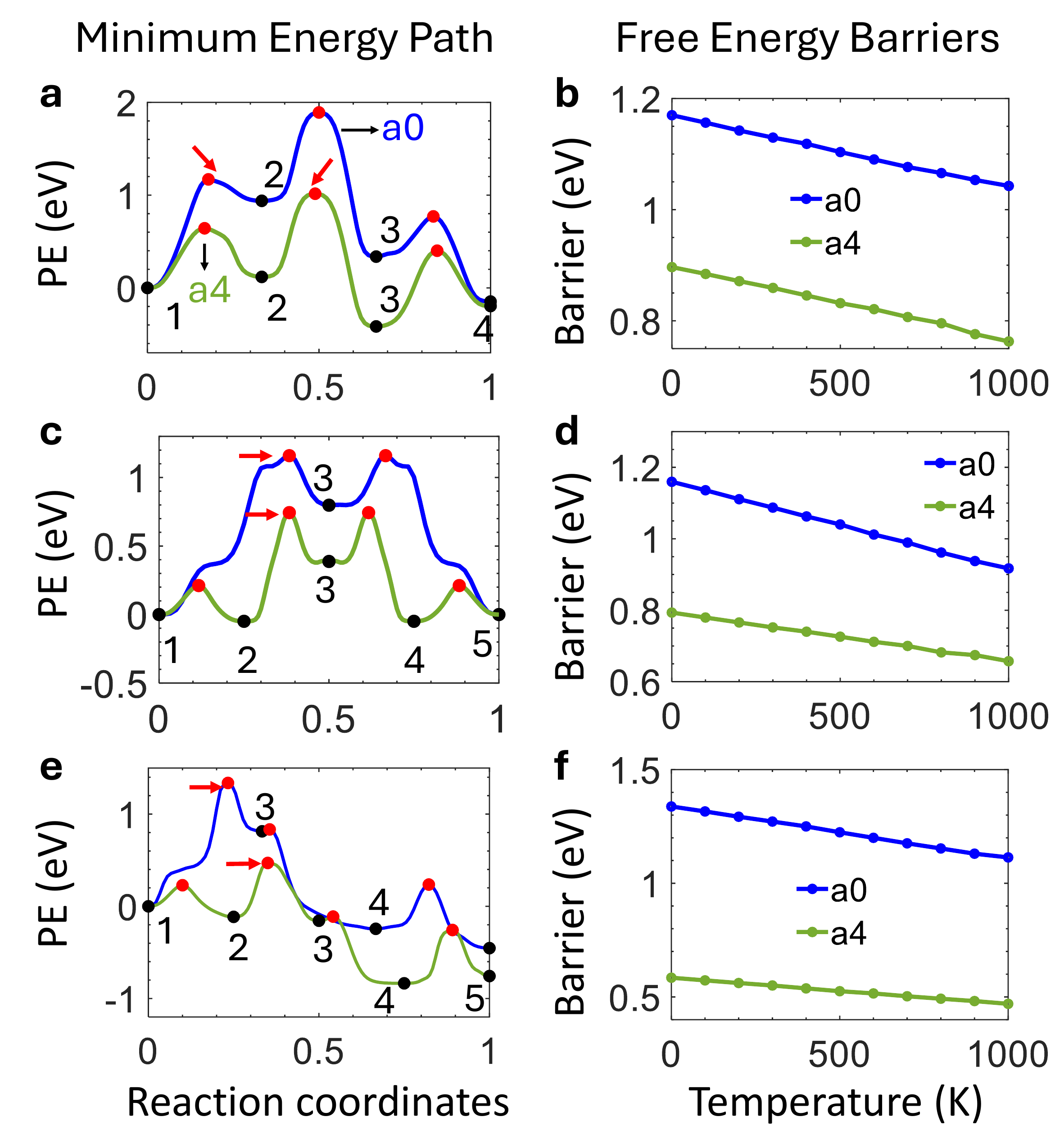}
    \caption{\textbf{Minimum‑energy pathways and temperature‑dependent free‑energy barriers for key events during 1T$^\prime$ nucleation under zero strain (a$_0$) and 4\,\% armchair tension (a$_4$).}
    \textbf{a,\,b}, Coalescence of Te monovacancies into a divacancy.  
    \textbf{c,\,d}, Migration of a divacancy through the lattice.  
    \textbf{e,\,f}, Nucleation of a triangular 1T′ island.}
    \label{figs-three-strain-pafi}
\end{figure}

\newpage

\begin{figure}[!ht]
    \centering
    \includegraphics[width=1\linewidth]{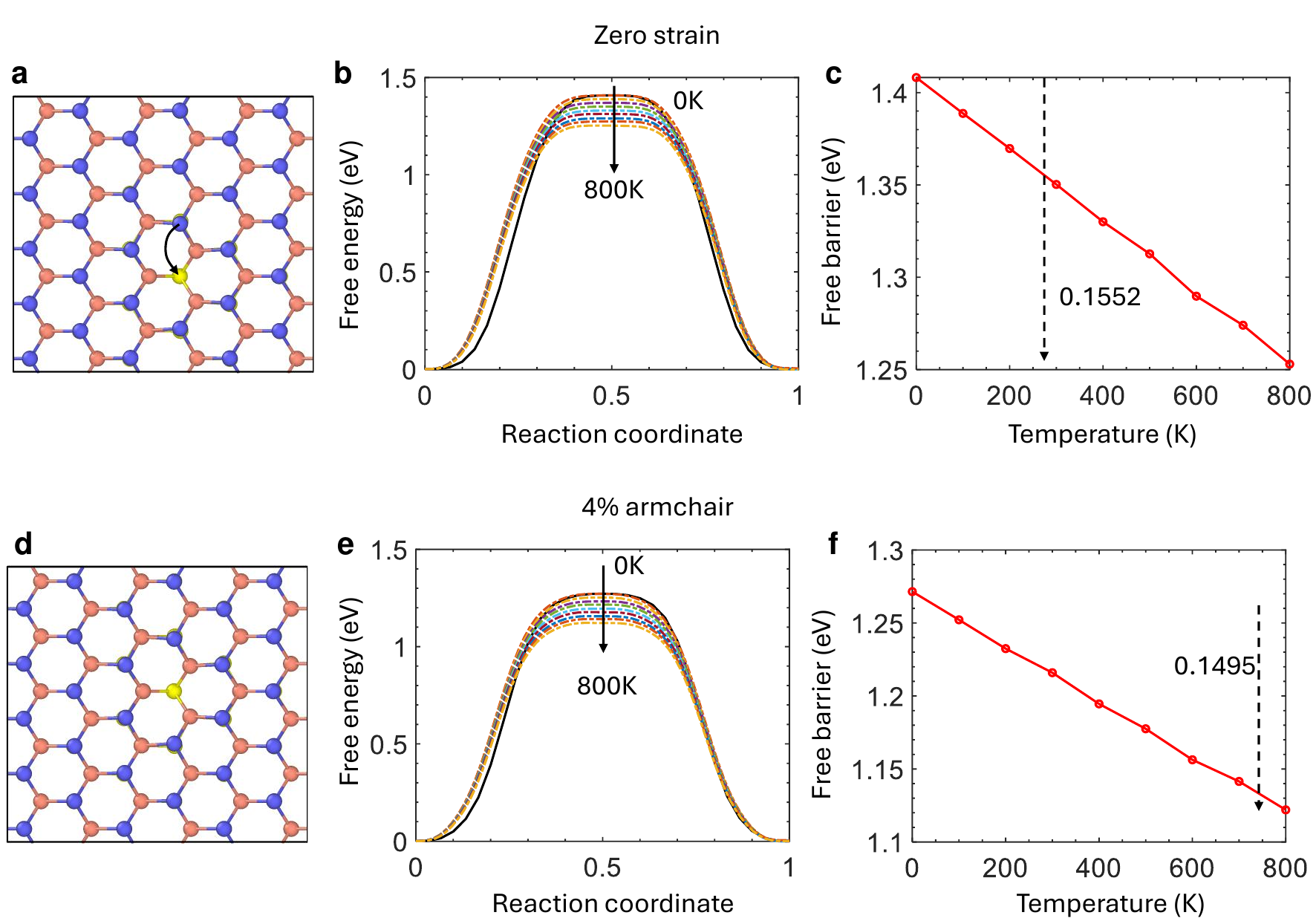}
    \caption{\textbf{Te monovacancy migration under varying temperature and strain conditions.}  
    \textbf{a, d} Structural evolution diagrams showing initial configuration, migration pathway (black arrow), and final relaxed structure.  
    \textbf{b, e} Minimum free energy paths as a function of temperature under (b) strain-free and (e) 4\% armchair-oriented strain.  
    \textbf{c, f} Corresponding migration energy barriers quantified across temperatures for (c) unstrained and (f) strained systems.}  
    \label{figs-MV-pafi}
\end{figure}
\newpage

\newpage

\begin{figure}[!ht]
    \centering
    \includegraphics[width=1\linewidth]{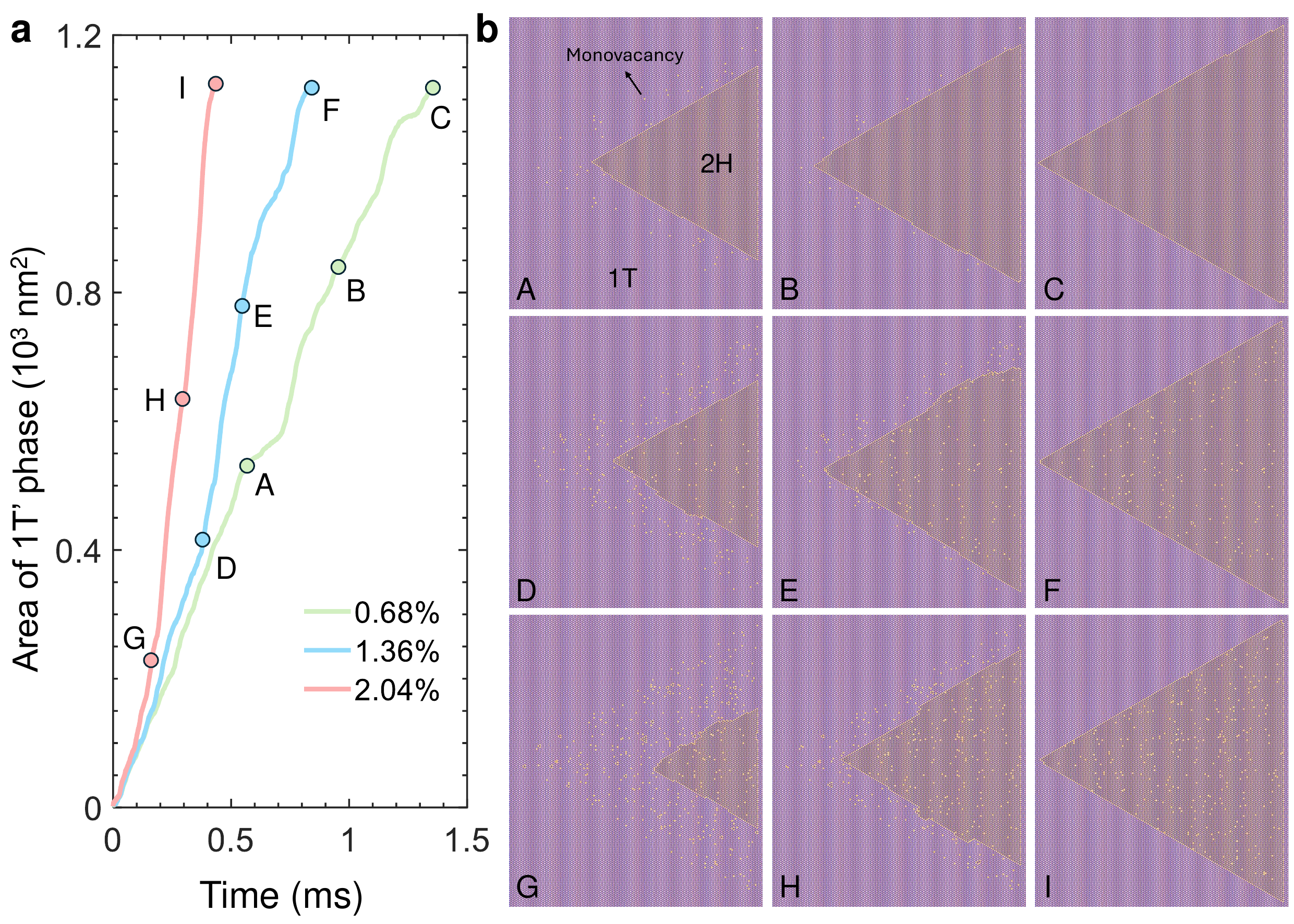}
    \caption{\textbf{KMC simulations of 1T$^\prime$ phase growth dynamics with varied Te vacancy concentrations.}  
    \textbf{a} Temporal evolution of the 1T$'$ phase area.  
    \textbf{b} Morphological development during growth.}  
    \label{figs-KMC-initial}
\end{figure}

\newpage

\begin{figure}[!ht]
    \centering
    \includegraphics[width=1\linewidth]{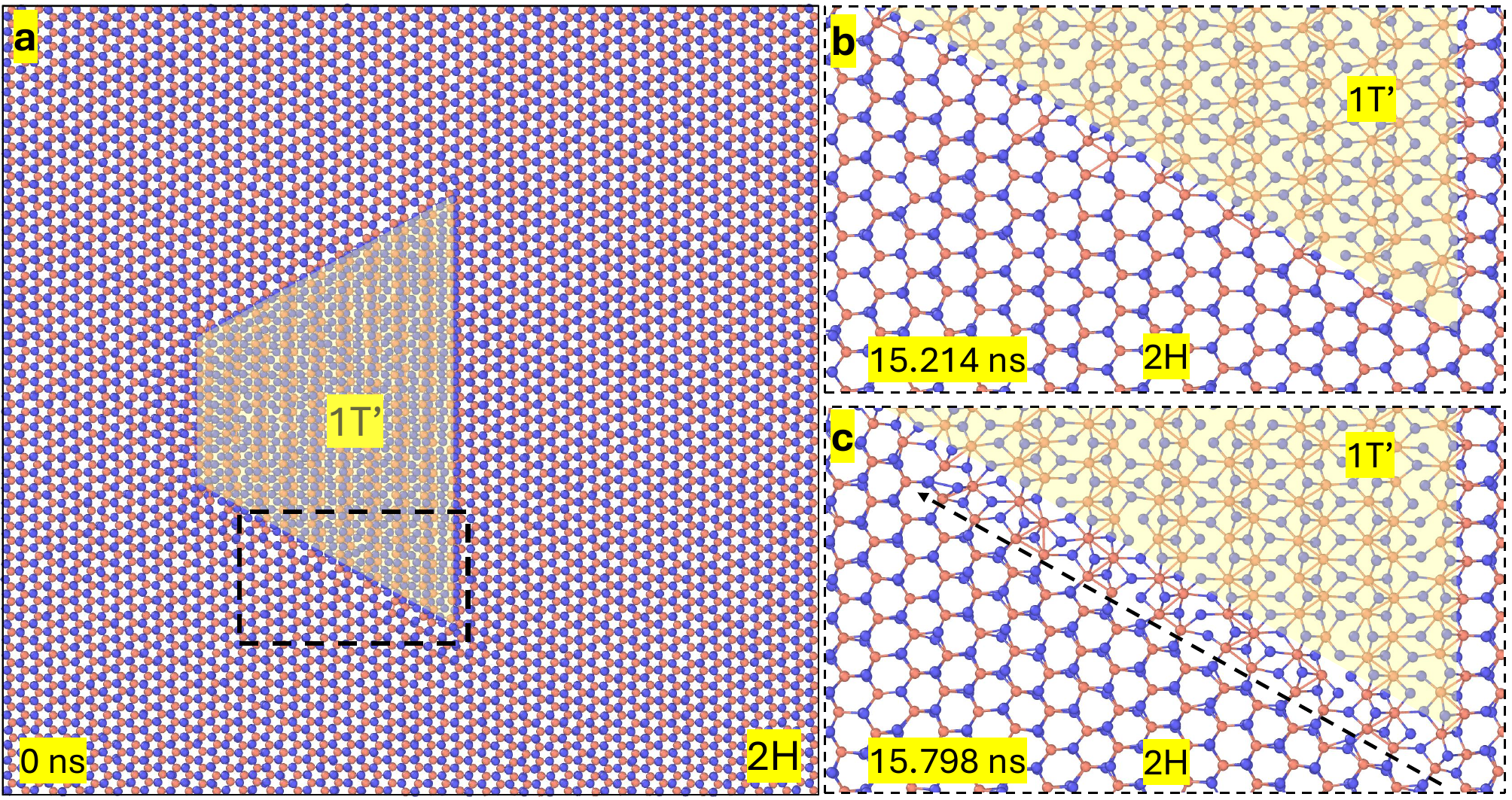}
    \caption{\textbf{Time‑resolved MD snapshots of defect‑free 1T$^\prime$ domain growth.} 
    \textbf{a}, Initial configuration: a single 1T$^\prime$ island embedded in a 2H matrix. 
    \textbf{b}, No discernible evolution after 15.214\,ns. 
    \textbf{c}, A sudden burst of 1T$^\prime$ expansion (along the dashed arrow) is observed at 15.798\,ns.}
    \label{figs-md-no-defect}
\end{figure}

\newpage

\begin{figure}[!ht]
    \centering
    \includegraphics[width=1\linewidth]{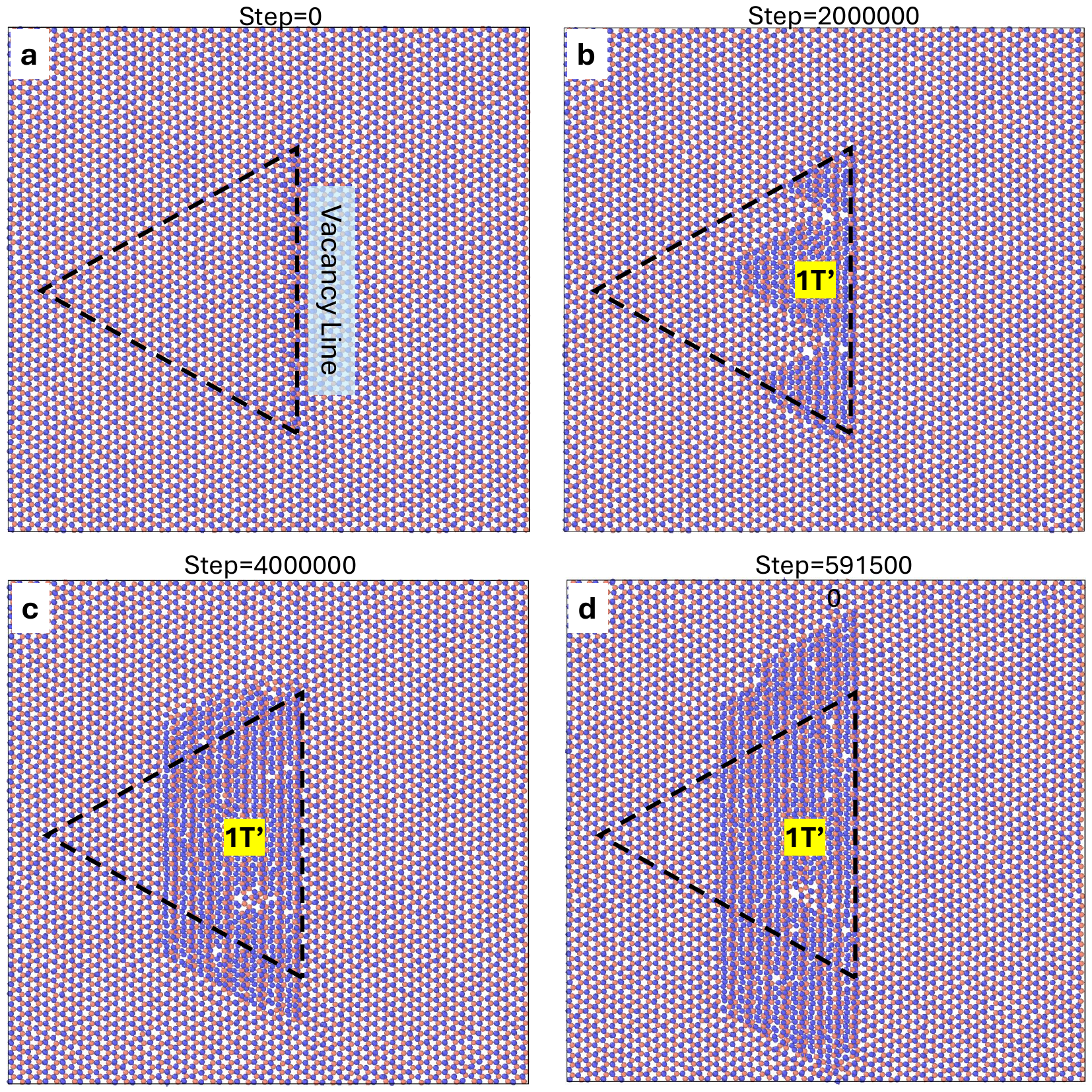}
    \caption{\textbf{tfMC simulations of 1T$^\prime$ phase domain growth from an extended Te vacancy line (VL).}  
    Structural evolution across sequential simulation stages, where dashed triangular outlines denote 1T$^\prime$ domains nucleated through VL-initiated low-energy pathways. The extended 1T$^\prime$ phase beyond these triangular regions in \textbf{c, d} emerges through vacancy-independent growth mechanisms detailed in Fig.~\ref{fig:reverse_transition}a.}
    \label{figs-tfmc-no-defect}
\end{figure}

\newpage

\begin{figure}[!ht]
    \centering
    \includegraphics[width=1\linewidth]{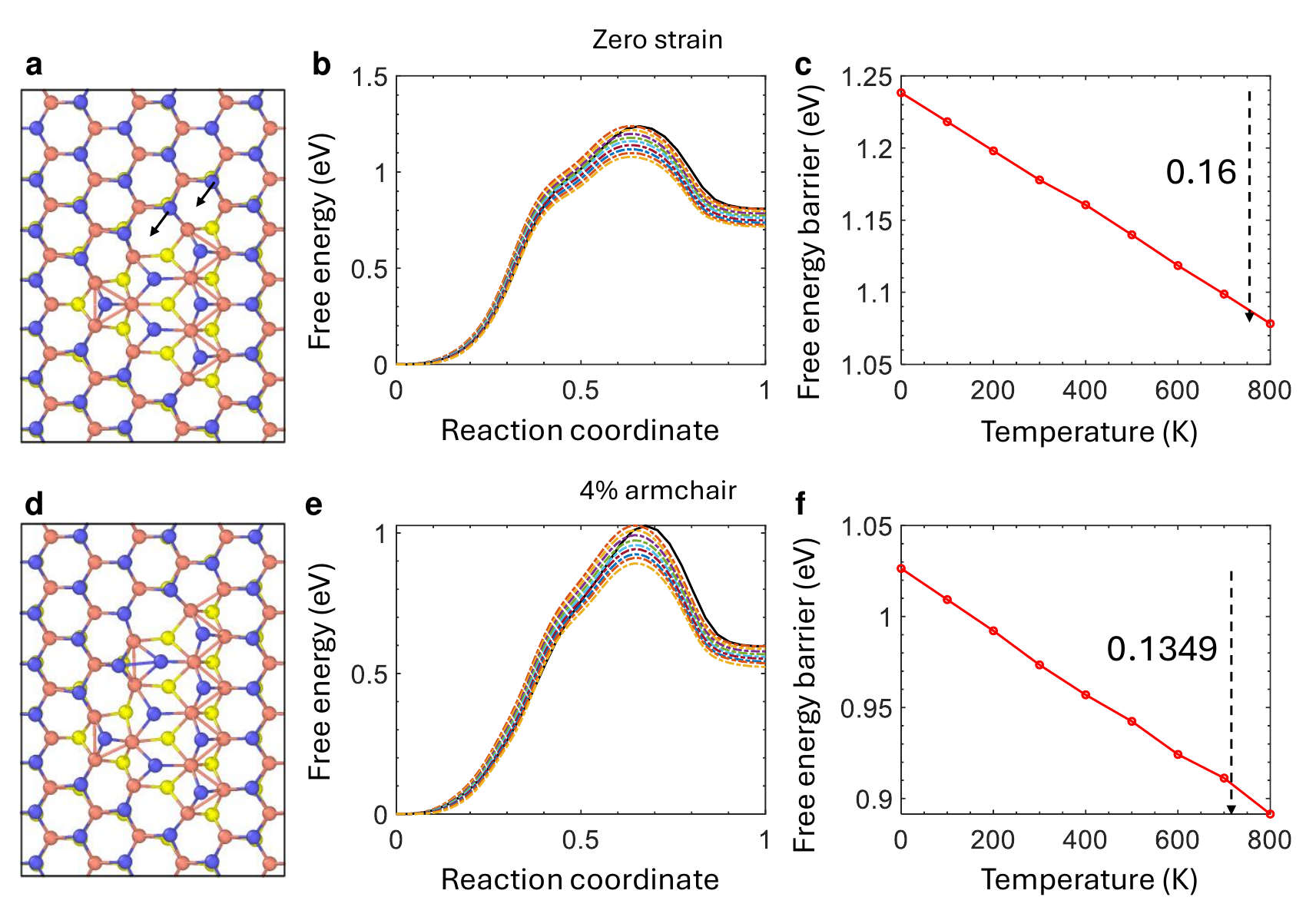}
    \caption{\textbf{Free energy barriers for 1T$^\prime$ phase growth in vacancy-free systems under strain.}  
    \textbf{a, d} Initial configuration, migration pathway (black arrows), and final 1T$^\prime$ phase.  
    \textbf{b, e} Minimum free energy paths across temperatures under (b) zero strain and (e) 4\% armchair-oriented strain.  
    \textbf{c, f} Temperature-dependent free energy barriers for phase transition under (c) strain-free and (f) 4\% armchair-strained conditions.}  
    \label{figs-pafi-no-defect}
\end{figure}

\newpage

\begin{figure}[!ht]
    \centering
    \includegraphics[width=1\linewidth]{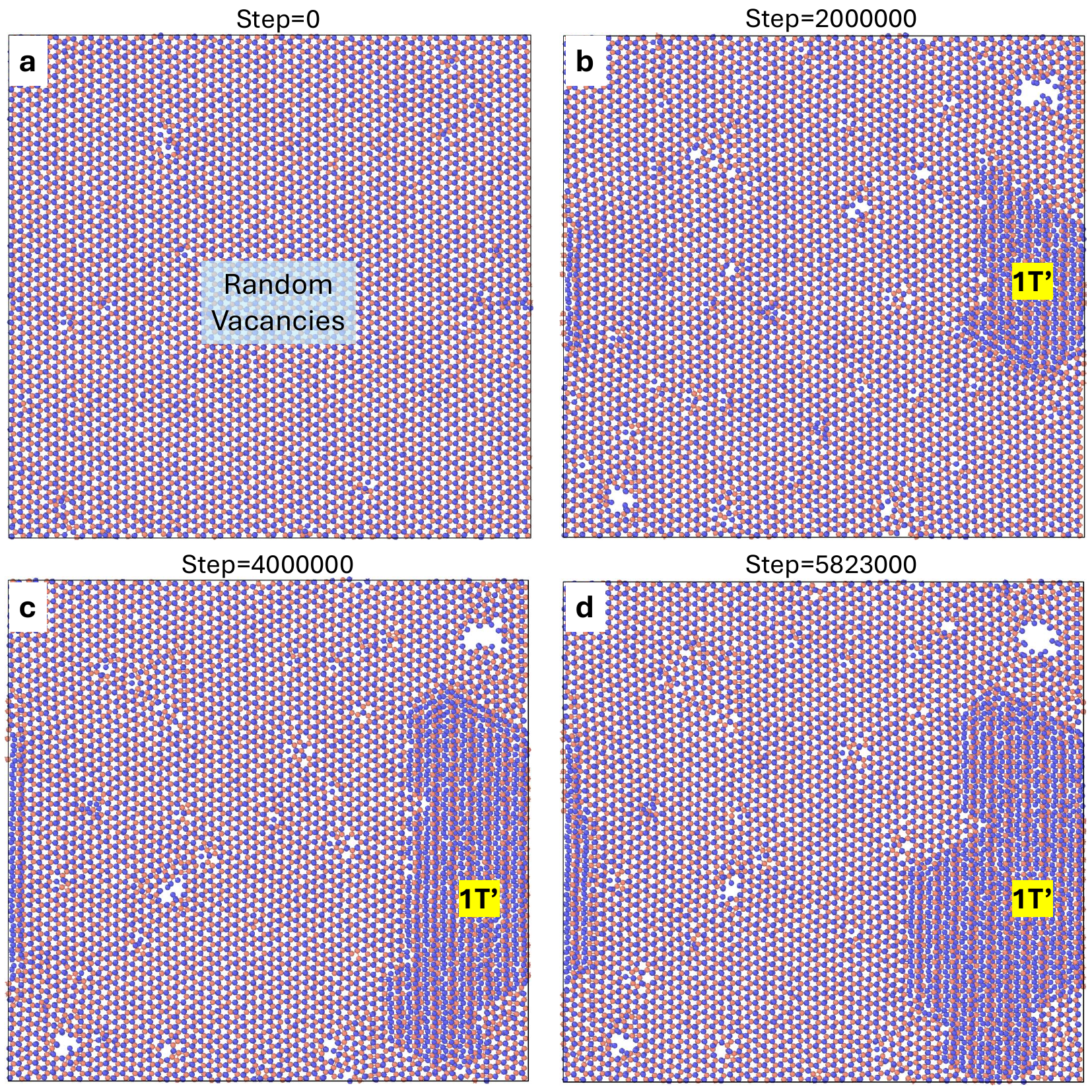}
    \caption{\textbf{tfMC simulation results of 1T$^\prime$ phase domain growth starting from random Te vacancies at different simulation steps.}}
    \label{figs-tfmc-random-vac}
\end{figure}

\newpage

\clearpage
\begin{table}[!htbp]
    \centering
    \caption{Characteristic time scale of critical events under 4\% armchair strain (MV: Te monovacancy; DV: Te divacancy).}
    \begin{tabular}{cccccc}
        \toprule
         \textbf{Events} &  {$E_{b,\text{300K}}$} (\si{eV}) & {$t_\text{300K}$ (\si{\second})} & {$t_\text{400K}$ (\si{\milli\second})} & {$t_\text{500K}$ (\si{\micro\second})} & {$t_\text{600K}$ (\si{\nano\second})} \\
         \midrule
         MV migration & 1.22 & $2.67\times10^{7}$ & $1.12\times10^{5}$ & $7.40\times10^{4}$ & $5.16\times10^{5}$ \\
         MV coalescence & 0.86 & 27.11 & 4.50 & 24.19 & 770 \\
         DV migration & 0.75 & 0.43 & 0.21 & 2.07  & 95 \\
         1T$^\prime$ growth without MV & 0.97 & $2.25\times10^{3}$ & 114.17 & 315.62 & $5.80\times10^{3}$ \\
         1T$^\prime$ growth with MV & 0.4 & $5.24\times10^{-7}$ & $1.10\times10^{-5}$ & $1.08\times10^{-3}$ & 0.23\\
         \bottomrule
    \end{tabular}
    \label{tab:time_scale}
\end{table}
\clearpage
\end{document}